\let\TeXyear\year
\documentclass[10pt]{llncs}

\let\year\TeXyear

\usepackage{cite}
\usepackage{amsmath}
\usepackage{amssymb}
\usepackage{amsfonts}
\usepackage{algorithmicx}
\usepackage{graphicx}
\usepackage{textcomp}

\usepackage{paralist}
\usepackage[usenames,dvipsnames]{xcolor}
\usepackage{stmaryrd} %\llbracket for \sem
\usepackage{algorithm}
\usepackage[noend]{algpseudocode}
\usepackage{listings}
\usepackage{cite}

\usepackage{macros/ltl}
\usepackage{macros/felipe-macros}
\usepackage{luismi}

\newcommand{\lola}{Lola\xspace}
\newcommand{\DIST}[2]{\ensuremath{\mathit{dist_{#1\_#2}}}}

\newcommand{\QUEUE}{\ensuremath{Q}}
\newcommand{\REQ}{\ensuremath{\mathbf{req}}}
\newcommand{\CONFIRM}{\ensuremath{\mathbf{confirm}}}
\newcommand{\RESP}{\ensuremath{\mathbf{resp}}}
\newcommand{\PEN}{\ensuremath{P}}
\newcommand{\WAIT}{\ensuremath{W}}

\newcommand{\SRC}{\ensuremath{\mathit{src}}}
\newcommand{\DST}{\ensuremath{\mathit{dst}}}
\newcommand{\TYPE}{\ensuremath{\mathit{type}}}
\newcommand{\STREAM}{\ensuremath{\mathit{stream}}}
\newcommand{\VAL}{\ensuremath{\mathit{val}}}
\newcommand{\MSG}{\ensuremath{\mathit{msg}}}

\newcommand{\READ}{\ensuremath{\mathit{read}}}
\newcommand{\INPUTS}{\ensuremath{\mathit{ins}}}
\newcommand{\OUTPUTS}{\ensuremath{\mathit{outs}}}

\newcommand{\Rold}{\ensuremath{R_{\mathit{old}}}}
\newcommand{\Rnew}{\ensuremath{R_{\mathit{new}}}}
\newcommand{\ISEAGER}{\ensuremath{\mathit{eager}}}
\newcommand{\ISLAZY}{\ensuremath{\mathit{lazy}}}

\newcommand{\SEND}{\ensuremath{\mathit{send}}}
\newcommand{\subf}{\ensuremath{\mathit{sub}}}
\newcommand{\APPEARS}{\ensuremath{\mathit{appears}}}
\newcommand{\DONE}{\ensuremath{\mathit{done}}}
\newcommand{\kNOT}{\ensuremath{\mathit{not}}}
\newcommand{\ISGROUND}{\ensuremath{\mathit{ground}}}
\newcommand{\Evaluate}{\ensuremath{\textsc{Evaluate}}\xspace}
\newcommand{\Subst}{\ensuremath{\textsc{Subst}}}

\newcommand{\OUTPUT}{\ensuremath{\textit{out}}}
\newcommand \false {\mathit{false}}
\newcommand \true {\mathit{true}}

\newcommand{\Osem}[1]{\ensuremath{[#1]}}

\newcommand{\mysc}[1]{\textnormal{\textsc{#1}}}
\newcommand{\Step}{\mysc{Step}\xspace}

\newcommand{\Finalize}{\mysc{Finalize}\xspace}
\newcommand{\Monitor}{\mysc{Monitor}\xspace}

\newcommand{\Prune}{\mysc{Prune}\xspace}

\newcommand{\ProcessMessages}{\mysc{ProcessMessages}\xspace}

\newcommand{\SendResponses}{\mysc{SendResponses}\xspace}

\newcommand{\SendRequests}{\mysc{SendRequests}\xspace}

\newcommand{\Now}{\mysc{Now()}\xspace}

\newcommand{\MTR}{\ensuremath{\textit{MTR}}\xspace} %Moment to Resolve
\newcommand{\MTRToRem}[1]{\ensuremath{#1_{\textit{rem}}}\xspace}
\newcommand{\MTRlazy}{\ensuremath{\MTR^{\textit{lazy}}}\xspace}
\newcommand{\MTRsyn}{\ensuremath{\MTR^{\textit{sync}}}\xspace} %Moment to Resolve
\newcommand{\MTRsynlazy}{\ensuremath{\MTR^{\textit{sync lazy}}}\xspace}
\newcommand{\MTRg}{\ensuremath{\MTR^{\mathbf{\globalTimeNameShort}}}\xspace} %Moment to Resolve
\newcommand{\MTRglazy}{\ensuremath{\MTR^{\globalTime~\textit{lazy}}}\xspace}
\newcommand{\MTRlocal}{\ensuremath{\textit{MTR}^{\textbf{\localTimeNameShort}}}\xspace} %Moment to Resolve
\newcommand{\MTRlocallazy}{\ensuremath{\MTR^{\localTimeShort~\textit{lazy}}}\xspace}

\newcommand{\TTR}{\ensuremath{\textit{TTR}}\xspace} %Time to Resolve

\newcommand{\TTRsyn}{\ensuremath{\textit{TTR}^{\textit{sync}}}\xspace} %Time to Resolve

\newcommand{\Term}{\ensuremath{\mathit{Term}}}

\newcommand{\simp}{\ensuremath{\mathit{simp}}}

\newcommand{\Arr}{\ensuremath{\mathit{arr}}\xspace}
\newcommand{\ArrivalTime}[2]{\Arr_{#1\to#2}\xspace}

\newcommand{\MonState}{\ensuremath{\mathit{MS}}\xspace}

\newcommand{\Spec}{\ensuremath{\varphi}\xspace}

\newcommand{\InstVar}[2]{\ensuremath{#1\tuple{#2}}\xspace}
\newcommand{\StreamVar}[1]{\ensuremath{#1}\xspace}

\algnewcommand\algorithmicswitch{\textbf{switch}}
\algnewcommand\algorithmiccase{\textbf{case}}
\algnewcommand\algorithmicassert{\texttt{assert}}
\algnewcommand\Assert[1]{\State \algorithmicassert(#1)}%
\algdef{SE}[SWITCH]{Switch}{EndSwitch}[1]{\algorithmicswitch\ #1\ \algorithmicdo}{\algorithmicend\ \algorithmicswitch}%
\algdef{SE}[CASE]{Case}{EndCase}[1]{\algorithmiccase\ #1}{\algorithmicend\ \algorithmiccase}%
\algtext*{EndSwitch}%
\algtext*{EndCase}%

\newcommand{\remove}{\mathit{del}}
\newcommand{\add}{\mathit{add}}
\newcommand{\IS}[3]{\InstVar{#1}{#2}\!\mapsto{}\!#3}

\newcommand{\depWeightSub}[4]{\ensuremath{#1\xrightarrow[#3]{#2}#4}\xspace}

\newcommand{\depWeightSubStar}[4]{\ensuremath{#1\xrightarrow[#3]{#2}^{*}#4}\xspace}

\newcommand{\TAmsgTrip}{\ensuremath{d}\xspace}

\newcommand{\EvGraph}{\ensuremath{\textit{EG}}\xspace}

\newcommand{\globalTimeName}{aeternal}
\newcommand{\globalTimeNameShort}{g}
\newcommand{\globalTime}{\textbf{\globalTimeName}\xspace}
\newcommand{\localTime}{\textbf{temporary}\xspace}
\newcommand{\globalTimeAdv}{\textbf{aeternally}\xspace}

\newcommand{\localTimeNameShort}{temp}
\newcommand{\localTimeShort}{\textbf{\localTimeNameShort}\xspace}
\newcommand{\GlobalTimeAdv}{\textbf{AETERNALLY}\xspace}
\newcommand{\LocalTimeAdv}{\textbf{TEMPORARILY}\xspace}
\newcommand{\Win}{\ensuremath{\textit{win}}\xspace}
\newcommand{\wdirrem}[3]{\depWeightSub{#1}{#2}{drem}{#3}\xspace}

\newcommand{\wloc}[3]{\depWeightSub{#1}{#2}{loc}{#3}\xspace}
\newcommand{\wlocnext}[2]{\xrightarrow[loc]{#1}{#2}}
\newcommand{\wremnext}[2]{\xrightarrow[rem]{#1}{#2}}
\newcommand{\dworst}{\ensuremath{d_{\textit{worst}}}}

\newcommand{\ACC}{\ensuremath{\mathit{acc}}}
\newcommand{\ROOT}{\ensuremath{\mathit{root}}}
\newcommand{\RESET}{\ensuremath{\mathit{reset}}}
\newcommand{\tconf}{\textit{tconf}\xspace}

\definecolor{StriverBlue}{RGB}{42, 0, 255}
\definecolor{StriverGreen}{RGB}{0, 155, 0}
\definecolor{StriverRed}{RGB}{155, 0, 0}
\definecolor{StriverOrange}{RGB}{255, 183, 68}
\definecolor{StriverYellow}{RGB}{155, 155, 45}
\lstdefinelanguage{Striver}
{
    basicstyle=\ttfamily,
    keywordstyle=[1]\color{StriverYellow},
    keywordstyle=[2]\color{StriverBlue},
    keywordstyle=[3]\bfseries,
    keywordstyle=[4]\color{StriverBlue},
    keywordstyle=[5]\color{StriverOrange},
    keywordstyle=[6]\color{StriverGreen},
    keywordstyle=[7]\color{StriverRed},
    otherkeywords = { <~,<<,~>,>>,macro,:=,const,input,output,true,false,ticks,define,outside,notick,U,delay,||,&&,+,-,==,!,!=,if,let,in then,else,int,bool,unit,Time },
    morekeywords = [1]{<~,<<,~>,>>},
    morekeywords = [2]{macro, :=, input,const, output, ticks, define},
    morekeywords = [3]{outside, notick, true, false},
    morekeywords = [4]{if, then, else},
    morekeywords = [5]{int, bool, unit, Time},
    morekeywords = [6]{U, delay, ||, &&, +, -, !, !=, ==},
    morekeywords = [7]{let, in},
}
\lstdefinelanguage{LOLA}
{
    basicstyle=\ttfamily,
    keywordstyle=[1]\color{StriverYellow},
    keywordstyle=[2]\color{StriverBlue},
    keywordstyle=[3]\bfseries,
    keywordstyle=[4]\color{StriverBlue},
    keywordstyle=[5]\color{StriverOrange},
    keywordstyle=[6]\color{StriverGreen},
    keywordstyle=[7]\color{StriverRed},
    otherkeywords = {macro,:=,const,input,output,true,false,define,outside,U,delay,||,&&,+,-,==,!,!=,if,let,in then,else,int,bool,unit,Time,num},
    morekeywords = [1]{},
    morekeywords = [2]{macro, :=, input, const, output,  define},
    morekeywords = [3]{true, false},
    morekeywords = [4]{if, then, else},
    morekeywords = [5]{int, bool, unit, num},
    morekeywords = [6]{U, delay, ||, &&, +, -, !, !=, ==, or, and, not, implies},
    morekeywords = [7]{let, in},
}

\newcommand{\AlgorithmTADLola}{
\begin{algorithm}[th!]
  \caption{Local monitoring algorithm at node $n$ with $\MonState_n = \tupleof{Q_n,U_n,R_n,\PEN_n,\WAIT_n}$}
  \label{alg:local-algo-tadLola}
  \begin{algorithmic}[1]
    \Procedure{\Monitor}{}
    \State $\MonState_n \gets \emptyset$; $k\gets \Now$ %gets clock or now()
    \While{\text{not END}} $\Step(k)$
    \EndWhile
    \State $M\gets k$; $\Finalize(M)$ %\Comment{Trace length $M$}
    \EndProcedure
	\vspace{0.3em}
    \Procedure{\Step}{$k$}
    \State $\Rold\gets \MonState_n.R_n$ %and U Pen, Waiting make it structured
    \State $\ProcessMessages(\MonState_n)$ %\Comment{Incoming messages}
    \State $R_n.\add(\{\IS{r}{k}{\READ(r,k)} \;|\; r\in\INPUTS_n \})$ %\Comment{Read inputs}
    \State $U_n.\add(\{\IS{s}{k}{\InstVar{e_s}{k}}\hspace{1.98em}\;|\;s\in\OUTPUTS_n\})$ %\Comment{Instantiate outputs}
    \State $\Evaluate(\MonState_n)$
    \State $\SendResponses(\MonState_n)$
    \State $\SendRequests(\MonState_n)$
    \State $\Prune(\MonState_n)$
    \EndProcedure
    \vspace{0.3em}
    \Procedure{\Evaluate}{$\MonState_n$}
    \State $\DONE \gets \false$
    \While {$\kNOT\;\; \DONE$}
    \State $\DONE \gets \true$
    \ForAll {$\IS{s}{k}{e} \in U_n$}
    \State $e'\gets \Subst(e,R_n)$
    \If {$\ISGROUND(e')$} $\DONE\gets \false$
    \State $U_n.\remove(\IS{s}{k}{e}) ; R_n.\add(\IS{s}{k}{e'})$
    \Else\ $U_n.\remove(\IS{s}{k}{e}); U_n.\add(\IS{s}{k}{e'})$
    \EndIf
    \EndFor
    \EndWhile
    \EndProcedure
	\vspace{0.3em}
    \Procedure{\ProcessMessages}{$\MonState_n$}
    \ForAll {$\MSG \gets Q_n.pop()$} 
    \Switch{$\MSG$}
    \Case{$\tuple{\REQ, \InstVar{s}{k}}$} \hspace{1.455em}$\PEN_n.\add(\InstVar{s}{k})$
    \EndCase
    \Case{$\tuple{\RESP, \InstVar{s}{k}, c}$}
    \State $R_n.\add(\IS{s}{k}{c}); W_n.\remove(\InstVar{s}{k})$
    \EndCase
    \EndSwitch
    \EndFor
    \EndProcedure
	\vspace{0.3em}
    \Procedure{\SendResponses}{$\MonState_n, \Rold$}
    \State $\Rnew \gets \MonState.R_n \ \Rold$
    \ForAll {$\IS{u}{k}{c} \in \Rnew$} \Comment{Eager new knowledge}
    \State $\SEND(\RESP,\InstVar{s}{k},c,n,n_r)$
    \EndFor
    \ForAll {$\tuple{\REQ, \InstVar{s}{k}, n_r, n}\in\PEN_n$} \Comment{Lazy requests}
    \If {$\IS{s}{k}{c}\in R_n$}
    \State $\SEND(\RESP,\InstVar{s}{k},c,n,n_r)$
    \State $\PEN_n.\remove(\tuple{\REQ, \InstVar{s}{k}, n_r, n})$
    \EndIf
    \EndFor
    \EndProcedure
    \vspace{0.3em}
    \Procedure{\SendRequests}{$\MonState_n$}
    \ForAll {$(\underline{\phantom{a}},e)\in U_n$ }
    \ForAll {$\InstVar{u}{k'}\in\subf(e)$}
    \If {$\InstVar{u}{k'}\notin \WAIT_n \And \mu(u)\neq n$} %\Comment{Send needed new requests}
    \State $\SEND(\REQ,\InstVar{u}{k'},n,\mu(u)); \WAIT_n.\add(\InstVar{u}{k'})$
    \EndIf
    \EndFor
    \EndFor
    \EndProcedure   
    \vspace{0.3em}
    \Procedure{\Prune}{$\MonState_n$} \Comment{Remote instant variable $v$ is resolved and request would had arrived if sent}
    \Comment{If $\InstVar{u}{j}$ will not be needed}
    \ForAll {$\IS{u}{j}{c}\in R_n~|~k \geq j+2*d+\MTR(v)$} 
    \State $R_n.\remove(\IS{u}{j_i}{c_i})\}$ \Comment{Remove}
    \EndFor
    \EndProcedure

  \end{algorithmic}
\end{algorithm}
}

\newcommand{\AlgorithmTADLolaEager}{ %only Eager
\begin{algorithm}[th!]
  \caption{Local monitor at node $n$ with $\tupleof{Q_n,U_n,R_n}$}
  \label{alg:alg-tadLola-eager}
  \begin{algorithmic}[1]
    \Procedure{\Monitor}{}
    \State $Q_n \gets \emptyset$; $U_n \gets \emptyset$; $R_n \gets \emptyset$; $k\gets \Now$ %gets clock or now()
    \While{\text{not END}} $\Step(k)$
    \EndWhile
    \State $M\gets k$; $\Finalize(M)$ %\Comment{Trace length $M$}
    \EndProcedure
	\vspace{0.3em}
    \Procedure{\Step}{$k$}
    \State $\Rold\gets \MonState_n.R_n$ %and U Pen, Waiting make it structured
    \State $\ProcessMessages(\MonState_n)$ %\Comment{Incoming messages}
    \State $R_n.\add(\{\IS{r}{k}{\READ(r,k)} \;|\; r\in\INPUTS_n \})$ %\Comment{Read inputs}
    \State $U_n.\add(\{\IS{s}{k}{\InstVar{e_s}{k}}\hspace{1.98em}\;|\;s\in\OUTPUTS_n\})$ %\Comment{Instantiate outputs}
    \State $\Evaluate(\MonState_n)$
    \State $\Rnew \gets \MonState.R_n \setminus \Rold$
    \State $\SendResponses(\MonState_n)$
    \State $\Prune(\MonState_n)$
    \EndProcedure
    \vspace{0.3em}
    \Procedure{\Evaluate}{$\MonState_n$}
    \State $\DONE \gets \false$
    \While {$\kNOT\;\; \DONE$}
    \State $\DONE \gets \true$
    \ForAll {$\IS{s}{k}{e} \in U_n$}
    \State $e'\gets \Subst(e,R_n)$
    \If {$\ISGROUND(e')$} $\DONE\gets \false$
    \State $U_n.\remove(\IS{s}{k}{e}) ; R_n.\add(\IS{s}{k}{e'})$
    \Else\ $U_n.\remove(\IS{s}{k}{e}); U_n.\add(\IS{s}{k}{e'})$
    \EndIf
    \EndFor
    \EndWhile
    \EndProcedure
	\vspace{0.3em}
    \Procedure{\ProcessMessages}{$\MonState_n$}
    \ForAll {$\MSG = \tuple{\RESP, \InstVar{s}{k}, c} \gets Q_n.pop()$} 
    \State $R_n.\add(\IS{s}{k}{c})$%; W_n.\remove(\InstVar{s}{k})$
    \EndFor
    \EndProcedure
	\vspace{0.3em}
    \Procedure{\SendResponses}{$\MonState_n, \Rnew$}
    \ForAll {$\IS{u}{k}{c} \in \Rnew$} %\Comment{Eager new knowledge}
    \State $\SEND(\RESP,\InstVar{s}{k},c,n,n_r)$
    \EndFor
    \EndProcedure
    \vspace{0.3em}
    \Procedure{\Prune}{$\MonState_n, \Rnew$} %\Comment{Pruning $R_n$ at node $n$ at instant $k$}
    \ForAll {$\IS{u}{j}{c}$ s.t. $now \geq \MTR(\InstVar{u}{j})$} 
    \State $R_n.\remove(\IS{u}{j_i}{c_i})\}$ \Comment{Remove}
    \EndFor
    \EndProcedure

  \end{algorithmic}
\end{algorithm}
}

\newcommand{\TableTADlolaTTRAnalysis}{{\small\begin{figure}[b!]
\hspace{-1.5em}\begin{tabular}{|l|r|r|r|r|r|r|} \cline{2-7}
\multicolumn{1}{c|}{} & \multicolumn{3}{c|}{const} & \multicolumn{3}{c|}{peak} \\ \cline{2-7}
\multicolumn{1}{c|}{} & min &  med & max & min &  med & max  \\ \hline
smartP      & 4 & 4 & 4 & 4 & 4 & 102 \\ \hline %checked Nov 24
smartPShort & 4 & 4 & 4 & 4 & 4 & 102  \\ \hline %checked Nov 24
tomsk       & 2 & 2 & 2 & 2 & 2 & 100  \\ \hline %checked Nov 24 for tl=840
contextAct  & 2 & 2 & 2 & 2 & 2 & 100  \\ \hline %checked Nov 24
orange4H    & 2 & 2 & 2 & 2 & 2 & 100  \\ \hline
\multicolumn{7}{c}{\begin{tabular}{c}\\[-0.5em](a) Const and Peak network behaviors \end{tabular}}\\[1em] %\hline

\cline{2-7}
\multicolumn{1}{c|}{} & \multicolumn{3}{c|}{normal} & \multicolumn{3}{c|}{normalPeak} \\ \cline{2-7}
\multicolumn{1}{c|}{} & min &  med & max & min &  med & max  \\ \hline
smartP      & 3 & 10 & 18 & 4 & 10 & 105 \\ \hline %checked Nov 24
smartPShort & 6 & 11 & 18 & 7 & 12 & 105  \\ \hline %checked Nov 24
tomsk       & 1 & 5 & 10 & 1 & 5 & 100 \\ \hline %checked Nov 24 for tl=840
contextAct  & 1 &  5 & 9 & 1 & 5 & 100  \\ \hline %checked Nov 24
orange4H    & 1 & 5 & 10 & 1 & 5 & 100  \\ \hline
\multicolumn{7}{c}{\begin{tabular}{c}\\[-0.5em](b) Normal and NormalPeak network behaviors \end{tabular}}\\[1em] %\hline

\end{tabular}
\caption{TTR analysis of Tadlola for different network behaviors}
\label{fig:tadLola-ttr-analysis}
\end{figure}}
}

\usepackage{multirow}
\usepackage{wrapfig}
\usepackage{hyperref}
\graphicspath{{figures/}} %figures can be included using only the filename

\def\BibTeX{{\rm B\kern-.05em{\sc i\kern-.025em b}\kern-.08em
   T\kern-.1667em\lower.7ex\hbox{E}\kern-.125emX}}

\begin{document}
%\history{Date of publication xxxx 00, 0000, date of current version xxxx 00, 0000.}
%\doi{10.1109/ACCESS.2017.DOI}

\title{Decentralized Stream Runtime Verification for Timed Asynchronous Networks}
%for ieee
%\author{\uppercase{Luis Miguel Danielsson}\authorrefmark{1,2} and \uppercase{C\'esar S\'anchez}\authorrefmark{2}\IEEEmembership{Senior, IEEE}}
%\address[1]{Universidad Polit\'ecnica de Madrid, Madrid, Spain}
%\address[2]{IMDEA Software Institute, Madrid, Spain email: {lm.danielsson@imdea.org} email:{cesar.sanchez@imdea.org}}
%for llncs
\author{Luis Miguel Danielsson\inst{1,2} \and C\'esar S\'anchez\inst{1}}
\institute{IMDEA Software Institute, Spain
  \and
  Universidad Polit\'ecnica de Madrid (UPM), Spain\\
  \email{\{luismiguel.danielsson,cesar.sanchez\}@imdea.org}
}
%\tfootnote{This paragraph of the first footnote will contain support 
%information, including sponsor and financial support acknowledgment. For 
%example, ``This work was supported in part by the U.S. Department of 
%Commerce under Grant BS123456.''}

%\markboth
%{Author \headeretal: Preparation of Papers for IEEE TRANSACTIONS and JOURNALS}
%{Author \headeretal: Preparation of Papers for IEEE TRANSACTIONS and JOURNALS}
%\corresp{*Luis Miguel Danielsson, IMDEA Software Institute. email: lm.danielsson@imdea.org}
%for ieee
%\input{abstract}
%\begin{keywords}
%decentralized monitoring, distributed, runtime verification, stream runtime verification, time asynchronous networks
%Enter key words or phrases in alphabetical 
%order, separated by commas. For a list of suggested keywords, send a blank 
%e-mail to keywords@ieee.org or visit \underline
%{http://www.ieee.org/organizations/pubs/ani\_prod/keywrd98.txt}
%\end{keywords}

%\titlepgskip=-15pt

\maketitle
%next 3lines for llncs
\begin{keywords}
decentralized monitoring, distributed, runtime verification, stream runtime verification, time asynchronous networks
\end{keywords}
\begin{abstract}
  We study the problem of monitoring distributed systems where
  computers communicate using message passing and share an almost
  synchronized clock.
  This is a realistic scenario for networks where the speed of the
  monitoring is sufficiently slow (at the human scale) to permit
  efficient clock synchronization, where the clock deviations is small
  compared to the monitoring cycles.
  This is the case when monitoring human systems in wide area networks, the Internet
  or including large deployments.

  More concretely, we study how to monitor decentralized systems where
  monitors are expressed as stream runtime verification
  specifications, under a timed asynchronous network.
  Our monitors communicate using the network, where messages can take
  arbitrarily long but cannot be duplicated or lost.
  This communication setting is common in many cyber-physical systems
  like smart buildings and ambient living.
  Previous approaches to decentralized monitoring were limited to
  synchronous networks, which are not easily implemented in practice
  because of network failures.
  Even when networks failures are unusual, they can require several
  monitoring cycles to be repaired.

  In this work we propose a solution to the timed asynchronous
  monitoring problem and show that this problem generalizes the
  synchronous case.
  We study the specifications and conditions on the network behavior that
  allow the monitoring to take place with bounded resources, independently
  of the trace length.
  Finally, we report the results of an empirical evaluation of an
  implementation and verify the theoretical results in terms of
  effectiveness and efficiency.
\end{abstract}

\section{Introduction}
\label{sec:intro}
%
%What is the problem
%
% \summary{explain briefly DRSV, timed async model of computation, RV
% vs static verification, related work for spec languages, DSRV}
% \luismitext{now monitor is an abstraction of a non-empty set of
% streams that are computed at a network node}

We study the problem of decentralized runtime verification of stream
runtime verification (SRV) specifications under the timed asynchronous
model of computation.
In decentralized monitoring a specification is decomposed into a
network of monitors that communicate by exchanging messages.
These monitors cooperatively evaluate the specification against a
trace of input observations performed at distributed locations.
We present a solution to the decentralized monitoring problem under
the timed asynchronous model of computation---in which processes share
a sufficiently synchronized global clock but where messages can take
arbitrarily long to arrive.

\emph{Runtime verification} (RV) is a dynamic technique for software
quality assurance that consists of generating a monitor from a formal
specification.
This monitor then inspects a single trace of execution of the system
under analysis.
In contrast to static verification techniques, RV sacrifices
completeness to provide a readily usable formal method, that for
example can be easily combined with testing and debugging.
One of the problems that RV must handle is to generate monitors from a
specification.
Early approaches to RV specification languages were based on temporal
logics~\cite{havelund02synthesizing,eisner03reasoning,bauer11runtime},
regular expressions~\cite{sen03generating}, timed regular
expressions~\cite{TRE}, rules~\cite{barringer04rule}, or
rewriting~\cite{rosu05rewriting}.
Another approach to monitor specifications is Stream Runtime
Verification (SRV)---pioneered by Lola~\cite{dangelo05lola}---which
defines monitors by declaring equations that describe the dependencies
between output streams of results and input streams of observations.
SRV is a richer formalism than most RV solutions that goes beyond
Boolean verdicts (like in logical techniques) by allowing
specifications that compute richer verdicts as output.
Examples include counting events and other statistics, computation of
robustness values or generating explanations of the errors.
See~\cite{dangelo05lola,faymonville16stream,gorostiaga18striver,danielsson19decentralized,gorostiaga20unifying}
for examples illustrating the expressivity of SRV languages.

Another important aspect of runtime verification is the operational
execution of monitors: how to collect information and how to perform
the monitoring task.
We focus in this paper in \emph{online} monitoring where the
monitoring happens incrementally as the input trace is being observed.
In
\cite{bauer12decentralised,elhokayem17monitoring,danielsson19decentralized}
the authors consider a centralized specification which gets deployed
as network of distributed monitors connected via a synchronous
network, where the global synchronous clock is used both for
communication and periodic sampling.
Monitors exchange messages and cooperate to perform the global
monitoring task.
This problem is called \emph{decentralized monitoring}
(see~\cite{francalanza18decentralised}).
We study here the timed asynchronous networks or communication
together with periodic sampling of inputs, that is a synchronous
computation over an asynchronous network.
Our solution subsumes the previously available SRV solution for
synchronous computation and a synchronous reliable network studied
in~\cite{danielsson19decentralized}.
We call the more general problem studied in this paper the \emph{timed
  asynchronous decentralized monitoring problem}.
Our goal is to generate local monitors at each node that collaborate
to monitor the specification, distributing the computational load
while minimizing the network bandwidth and the latency of the
computation of verdicts.
Apart from more efficient evaluation, decentralized monitoring can
provide fault-tolerance as the process can partially evaluate a
specification using the information provided by the part of the
network that does not fail.
In the same spirit, if part of the network of cooperating monitors is
clogged---in the sense that it is working slower for some reason---the
other part can keep its normal throughput.
Consider for example an \textit{if-then-else} specification with a
slow computation needed to obtain the value of both the \textit{then}
and the \textit{else} parts.
Consider a decentralized deployment with three monitors connected as a
tree: the leaf monitors compute the \textit{then} and the \textit{else}
parts, while the root monitor computes the specification using a
Boolean input stream for \textit{if} part.
Assume that the condition is true $90\%$ of the time, so most of the
time the \textit{then} value is used and the \textit{else} value is
discarded.
Now, also consider that the network link between the root monitor and the leaf monitor that
computes the \textit{else} part is slow, the throughput of the root of the
specification remains unaffected for that $90\%$ of times and the
result can be produced without waiting for the long computation and
the network delay of the link that affect the \textit{else} part.
We plan to leverage the advantages of the decentralized systems to
aggressively incorporate fault-tolerance in future work.

\subsubsection{Our Solution.}

% \summary{explain an overview of the solution, tell the assumptions and details on the environment: network, topology...}

In this paper we provide a solution to the decentralized monitoring
problem for Lola~\cite{dangelo05lola} specifications for arbitrary
network topologies and placement of the local monitors.

We study \emph{time asynchronous} networks~\cite{cristian99timed},
where nodes share a global clock (built upon bounding the network
synchronicity delays and hardware clock drifts) but monitoring
messages can take arbitrarily long.
Time asynchronous networks~\cite{cristian99timed} ``... allow
practically needed services such as clock synchronization, membership,
consensus, election, and atomic broadcast to be implemented''.
%
% Synchronous specs
%
Synchronous networks are a special case where, additionally, messages
take a known bounded time to arrive.
We use the fact that a global clock is available to use a model of
computation for monitoring that proceeds in rounds, where each round
consists on input readings and process incoming messages, followed by
an update the internal state of local monitors and finally producing
output messages.
This synchronous execution model is realistic in many scenarios, for
example in smart buildings or smart cities---where clocks can be
synchronized using a time network protocol---that is sufficiently
precise for round cycles of tens of seconds.
We also assume in this paper a reliable system: nodes do not crash,
and messages are not lost or duplicated.
In our solution, different parts of the specification (modeled as
streams), including input readings, are deployed into different
network nodes as a local monitor.
Local monitors will communicate with other monitors when necessary to
resolve the streams assigned to them, trying to minimize the
communication overhead.
Intuitively, data will be read from sensor monitors, and then each
layer of intermediate monitors will compute sub-expressions and
communicate partial results to remote monitors in charge of
super-expressions, ultimately computing the stream of values of the
root expression.
A degenerated case of this setting is a centralized solution: nodes
with mapped observations send their sensed values to a fixed central
node that is responsible of computing the whole specification.
The SRV language that we consider is
Lola~\cite{dangelo05lola,sanchez18online}.
We will identify those specifications and conditions on the
network behavior that allow the monitoring to take place with bounded resources,
independently of the trace length.
\subsubsection{Motivating Example.}

\begin{example}
  We use as a running example a smart building with rooms equipped
  with sensors and a central node.
  The aim is to generate alarms when there is a fire risk.
  The following specification captures this risk by detecting acute
  uprisings in temperature and $\textit{CO}_2$ in a certain room.
  We place the computations needed to decide whether the measured
  variables rise `enough' to those nodes where the sensor readings
  take place.
  In this way, the central node only needs to compute which nodes
  present both the temperature and the $\textit{CO}_2$ alarm.
  We omit the $\textit{CO}_2$ computation for simplicity and readability (as it is an
  exact mirror of the temperature computation).
  The $\textit{CO}_2$ values would be useful to assess
  the risk of fire at the 'Building' monitor.
  
{\footnotesize\begin{lstlisting}[language=LOLA] 
@Room1{
  input num t_1   eval
  #tini_1 is a constant
  # with meaningful bounds
  define num tlow = 1.6 * tini_1
  define num thi  = 2.0 * tini_1
  define num t_spike_q_1 =
    if      t_1 <= low then 0
    else if t_1 >  hi  then 1
    else (t_1 - low)/(hi - low)
}
@Room2{
  input num t_2 eval
  define num t_low = 1.6 * tini_2
  define num t_hi  = 2.0 * tini_2
  define num t_spike_q_2 =
    if      t_2 <= t_low then 0
    else if t_2 >  t_hi then 1
    else (t_2 - t_low)/(t_hi-t_low)
}
@Building{
  define bool fire_risk_q_1 = t_spike_q_1 > 0.5
  define bool fire_risk_q_2 = t_spike_q_2 > 0.5
}
\end{lstlisting}}

\end{example}

\subsubsection{Related work.}
The term \emph{decentralized monitoring} is used in the
survey~\cite{francalanza18decentralised} to distinguish the term from
distributed monitoring where processes do not share a global clock.
In distributed monitoring a complete asynchronous network is assumed,
while typically decentralized monitoring assumes a completely
synchronous network where all samples and communication occur in
lockstep.
In this paper we explore the middle ground: network nodes share a
sufficiently synchronized global clock (like in synchronous
distributed systems) but communication can take arbitrarily long (like
in asynchronous distributed systems).
Also, in~\cite{francalanza18decentralised} they present other concepts
such as policy checking that are called decentralized monitoring that
do not correspond to the monitoring presented in this paper, because
they are concerned only about global safety properties that can be
used for asynchronous networks with asynchronous computations.

In~\cite{ganguly2021distributed} they also study timed asynchronous
networks of cooperating monitors but use an SMT-solver for simplifying
LTL formulas.

%\luismitext{search some of borzoo bonakdarpour, who studies consensus
%  and voting algorithms in purely distributed (async net async comp)
%  systems, recent work by borzoo is timed asynch: Crash-Resilient
%  Decentralized Synchronous Runtime Verification Shokoufeh Kazemlou;
%  Borzoo Bonakdarpour and others}.
%\luismitext{search some El-Hokayem and Yles Falcone papers on decentralized monitoring}

%  \summary{some other stream processing systems, related work for
%  distributed, decentralzied, other Lolas...}
%
Distributed stream processing has been largely studied.
In~\cite{dolequoc19PrivApprox} they use the concept of streams in
Complex Event Processing, where events may be structured datatypes and
where computation may be complex in the sense that several operations
are needed for each event, for example in sliding window operations to
make aggregate calculations on the arriving events.
The aim of~\cite{dolequoc19PrivApprox} is merging privacy and
approximation techniques obtaining zero-knowledge privacy and
low-latency and efficient analytics.
% , their queries are SQL queries and the clients answers are
% randomized, sometimes answers truthfully and others randomly, vector
% of boolean answers!!  In ~\cite{martin19lowcost}
%
In ~\cite{carbone2015apache} Apache Flink is introduced where stream
dataflows processing is used to handle continuous streams and batch
processing.
Distributed and decentralized monitoring has been studied in the
context of runtime verification.
Sen et al.~\cite{sen04efficient} introduces PT-DTL, a variant of LTL
logic for monitoring distributed systems, but they consider a complete
asynchronous distributed system and they are limited to Boolean
verdicts.
The work in~\cite{francalanza18decentralised} uses slices to support
node crashes and message errors when monitoring distributed message
passing systems with a global clock.
Bauer et al.~\cite{bauer13propositional} introduce a first-order
temporal logic and trace-length independent spawning automaton, and
in~\cite{bauer12decentralised} show a decentralized solution to
monitor $\textrm{LTL}_3$ in synchronous systems using formula
rewriting.
$\textrm{LTL}_3$ is a three-valued variant of LTL with a central value
in the lattice that captures when an expression has an unknown value
so far and we need to process more of the input trace to determine its
truth value.
This is improved in \cite{elhokayem17monitoring,elhokayem17themis}
using an Execution History Encoding (EHE).
EHE is a datastructure that stores the partially
evaluated expressions by different monitiors with their partial information
that allow decentralized monitors to infer the state in which the monitoring automaton is in.
In~\cite{monitoringDecentSpecs2020El-Hokayem} the authors extend the
EHE with distributed and multi-threaded support along with
guaranteeing the determinism of the datastructure by construction.
Then they analyze the compatibility and monitorability of
decentralized specifications using EHE.
However, the verdicts and data are still Boolean and the network
assumption is synchronizity.
In~\cite{globalChoreosSynthesis2020El-Hokayem} global choreographies
(as a kind of master-based protocol) are synthesized (including
control flows, synchronization, notification, acknowledgment,
computations embedding) to distributed systems.
Also, they provide a transformation to Promela which allows to verify
the implementation using LTL specifications.
Some schemes that they showcase are a variant producer-consumer or
two-phase commit and apply it to building micro-services such as a
buying system.
This work focuses on synthesizing the flow of monitors, but again the
observations and verdicts are Boolean.
In~\cite{crash-resilient2018Borzoo} a synchronous network of
LTL-monitors cooperate to achieve a verdict on the system under test
while they may suffer crashes.
In this scenario an SMT-based algorithm for synthesizing the automata
for the LTL-monitors is presented that achieves fault tolerance
providing soundness even though crashed monitors never recover.
Even though this work considers failures (which is out of the scope of
our paper) they assume synchronous communication.
All these approaches consider only Boolean verdicts.
In comparison, SRV can generate verdicts from arbitrary data domains.

All previous SRV efforts, from Lola~\cite{dangelo05lola},
Lola2.0~\cite{faymonville16stream},
Copilot~\cite{pike10copilot,pike13copilot,perez20copilot} and
extensions to timed event streams, like
\tessla~\cite{convent18tessla}, RTLola~\cite{faymonville19streamlab}
or \striver~\cite{gorostiaga18striver} assume a centralized monitoring
setting.
In \cite{gorostiaga20unifying} the relationship between time-based
(soft real time) and event-based models of computation and their
effects on SRV are explored, but again in the centralized setting.
The work in~\cite{basin15failure} shows how monitoring Metric Temporal
Logic specifications of distributed systems (including failures and
message reordering) where the nodes communicate in a tree fashion and
the root emits the final verdict.
The work in~\cite{danielsson19decentralized} proposes a solution to
the synchronous monitoring of SRV specifications but assuming a synchronous network.
We extend~\cite{danielsson19decentralized} to timed asynchronous networks.
\subsubsection{Contributions and structure.}
% \summary{contirb algorithm, resources analysis, implementation, empirical eval}
%
The main contribution of this paper is a solution, described in
Section~\ref{sec:solution}, to the timed asynchronous decentralized
stream runtime verification problem.
We provide a proof of correctness of the algorithms and show that our
solution subsumes a synchronous decentralized problem without
overhead.
A second contribution, included in Section~\ref{sec:resources-eager},
is the description of those specifications and conditions on the
network behavior that allow the monitoring to take place with bounded
resources, independently of the trace length.
Bounding resources is of the uttermost importance in cyber-physical
systems where memory, bandwidth and even computing time are limited in
order to react properly and timely to the changing environment.
If a cyber-physical system is trace-length independent it can run
indefinitely long even if the resources are physically constrained.
A third contribution, detailed in Section~\ref{sec:empirical}, is a
prototype implementation and an empirical evaluation.
A fourth contribution, in Section~\ref{sec:lazy}, is a modified
algorithm that allows nodes to save bandwidth by only communicating
stream values when requested.
Section~\ref{sec:lola} contains the preliminaries and
Section~\ref{sec:conclusions} concludes.
%
%Missing proofs appear in the appendix.

%%% Local Variables:
%%% TeX-master: "main.tex"
%%% TeX-PDF-mode: t
%%% End:

\section{Preliminaries. Stream Runtime Verification}
\label{sec:lola}
% \luismitext{the whole section is taken from self-stabilizing}
% \summary{explain lola in a nutshell}
We recall now SRV briefly.
For a more detailed description see~\cite{dangelo05lola} and the
tutorial~\cite{sanchez18online}.
The fundamental idea of SRV, pioneered by Lola~\cite{dangelo05lola} is
to describe monitors declaratively via a set of equations that
describe the dependencies between output streams of values and
input streams of values.
We focus here on online monitoring.
A monitor is generated from a specification, which at runtime computes
a sequence of values for the output streams as soon as possible after
observing each value from input streams.
Input values are typically extracted from some sensor or read from a
log file.

A Lola specification declares output streams in relation to the input
streams, including both future and past temporal dependencies.
The Lola language cleanly separates the temporal dependencies from the
individual operations to be performed at each step, which leads to
generalization of monitoring algorithms for logics to the computation
of richer values such as numbers, strings or richer data-types.

\subsection{Lola Syntax.}
%\summary{classic lola syntax}
A Lola specification consist of declaring the relation between output
streams and input streams of events.
Stream expressions are terms built using a collection of (interpreted)
constructor symbols.
Symbols are interpreted in the sense that each constructor is not only
used to build terms, but it is also associated with an evaluation
function, that given values of arguments produces a value of the
return type.
Given a set $Z$ of typed stream variables the set of \emph{stream
  expressions} consists of (1) variables from $Z$, (2) offsets
$v[k,d]$ where $v$ is a stream variable of type $D$, $k$ is a natural
number and $d$ a value from $D$, and (3) terms $f(t_1,\ldots,t_n)$
using constructor symbols $f$ from the theories to previously defined
terms.
Stream variables represent sequences of values (streams) in the
specification.
The intended meaning of an offset expression $v[-1,\false]$ is the
value of stream $v$ in the previous position of the trace (or $\false$
if there is no such previous position, that is, at the beginning).
We use $\Term_D(Z)$ for the set of stream expressions of type $D$
constructed from variables from $Z$ (and drop $Z$ if clear from the
context).
Given a term $t$, $\subf(t)$ represents the set of sub-terms of $t$.

\begin{definition}[Specification]
  A Lola specification $\varphi(I,O)$ consists of a set
  $I=\{r_1,\ldots,r_m\}$ of input stream variables, a set
  $O=\{s_1,\ldots,s_n\}$ of output stream variables, and a set of
  defining equations,
  \( s_i= e_i(r_1, \ldots, r_m, s_1, \ldots, s_n) \) 
  one per output variable $s_i\in O$.
  The term $e_i$ is from $\Term_D(I\cup O)$, where $D$ is the type of
  $s_i$.
\end{definition}
We will use $r$, $r_i$\ldots to refer to input stream variables; $s$,
$s_i$\ldots to refer to output stream variables; and $u$, $v$ for an
arbitrary input or output stream variable.
Given $\varphi(I,O)$ we use $\APPEARS(u)$ for the set of output
streams that use $u$, that is
$\{ s_i\;|\; u[-k,d]\in \subf(e_i) \textrm{ or } u\in\subf(e_i) \}$.
Also, $\ISGROUND(t)$ indicates whether expression $t$ is a ground
expression (contains no variables or offsets) and therefore can be
evaluated into a value using the interpretations of constants and
function symbols.

\begin{example}
  \label{ex:spec}
  The property ``\textit{sum the previous values in input stream
    \lstinline[language=LOLA]!y!, but if the \lstinline[language=LOLA]!reset! stream is true, reset the
    count}'', can be expressed as follows, where stream variable
  \texttt{\textup{root}} uses the accumulator \texttt{\textup{acc}}
  and the input \texttt{\textup{reset}} to compute the desired sum.
  The stream \texttt{\textup{acc}} is defined with the keyword
  \lstinline[language=LOLA]!define! to emphasize that it is an
  intermediate stream.
\end{example}
\vspace{-0.5em}

{\footnotesize\begin{lstlisting}[language=LOLA]
  input bool reset
  input num y 
  define int acc  = y + root[-1|0] 
  output int root = if reset then 0 else acc
\end{lstlisting}}

\subsection{Lola semantics.}

We introduce now the formal semantics of Lola, that guarantee that
there is a unique correct output stream for each input stream.
This semantics allows to prove that an algorithm is correct by showing
that the algorithm produces the desired output.
At runtime, input stream variables are associated incrementally with
input streams of values.

Given an input streams $\sigma_I$ (one sequence per input stream
variable) and given an output candidate $\sigma_O$ (one sequence per
output stream) the formal semantics captures whether the pair
$(\sigma_I,\sigma_O)$ matches the specification, which we write
$(\sigma_I,\sigma_O)\models\varphi$.
We use $\sigma_r$ for the stream in $\sigma_I$ corresponding to input
variable $r$ and $\sigma_r(k)$ for the value of stream $\sigma$ at
position $k$.
For $(\sigma_I,\sigma_O)\models\varphi$ to hold, all streams must be
sequences of the same length.

A \emph{valuation} of a specification $\varphi$ is a pair
$\sigma:(\sigma_I,\sigma_O)$ that contains one stream (of values of
the appropriate type) and of the same length for each input and output
stream variable in $\varphi$.
Given a term $t$, the \emph{evaluation} $\sem{t}_\sigma$ is a sequence
of values of the type of $t$ defined as follows:
\begin{itemize}
\item If $t$ is a stream variable $u$, then
  $\sem{u}_\sigma(j)=\sigma_u(j)$.
\item If $f=f(t_1,\ldots,t_k)$ then
  $\sem{f(t_1,\ldots,t_k)}_\sigma(j)=f(\sem{t_1}_\sigma(j),\ldots,\sem{t_k}_\sigma(j))$.
\item Finally, if $t=v[i,c]$ is an offset then
  $\sem{\InstVar{v}{j+i}}$ if $j+i$ is a valid point of the trace, and
  the defalult value $c$ otherwise.
  Formally: $\sem{v[i,c]}_\sigma(j)=\sem{v}_\sigma(j+i)$ if $0\leq j+i$, and $c$ otherwise.
\end{itemize}

A valuation $(\sigma_I,\sigma_O)$ satisfies a Lola specification
$\varphi$ whenever for every output variable $s_i$,
  \(
  \sem{s_i}_{(\sigma_I,\sigma_O)}=\sem{e_i}_{(\sigma_I,\sigma_O)}.
  \)
  In this case we say that $\sigma$ is an evaluation model of
  $\varphi$ and write $(\sigma_I,\sigma_O)\models\varphi$.

  The intention of a specification $\varphi$ is to describe a unique
  output from a given input, which is guaranteed if $\varphi$ has no
  cycles in the following sense.
  A \emph{dependency graph} $D_\varphi$ of a specification
  $\varphi(I\cup O)$ is a weighted multi-graph $(V,E)$ whose vertices
  are the stream variables $V=I\cup O$, and where $E$ contains a
  directed weighted edge $u\xrightarrow{w} v$ whenever $v[w,d]$ is a
  sub-term in the defining equation of $u$.
  A specification $\varphi$ is \emph{well-formed} if $D_\varphi$
  contains no zero-weight cycles, which guarantees that no stream
  depends on itself at the current position.

  Considering example~\ref{ex:spec}.
  Its dependency graph is:

  %\vspace{0.5em}
  %{\centering
  %  \hspace{8em}\includegraphics[scale=0.33]{figures/depGraphBlack}
  %}

  {\small\begin{figure}[tbh!]
  \centering
  \includegraphics[scale=0.3]{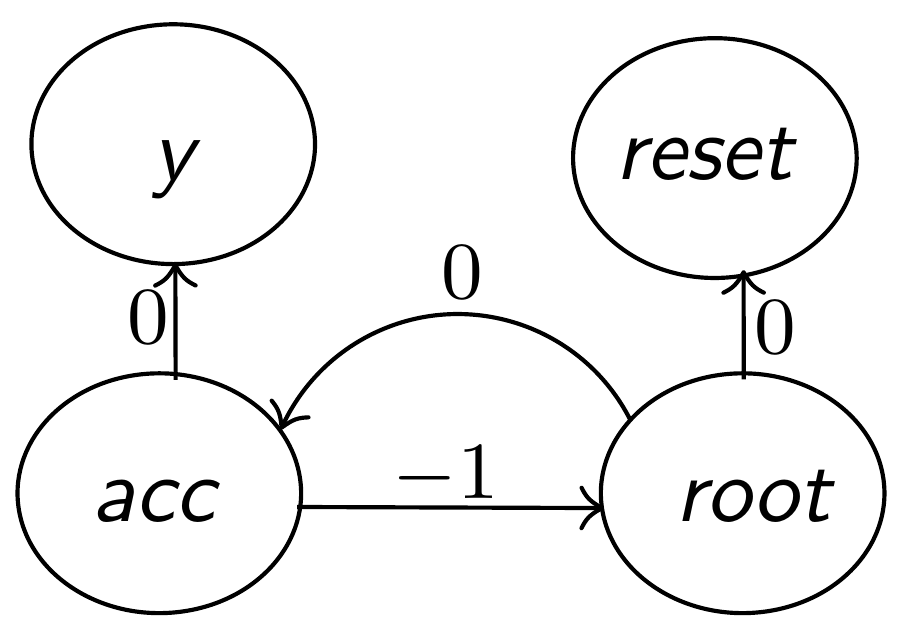} 
  \caption{Dependency graph for example~\ref{ex:spec}}
  \label{plot:evalGraph}
  \end{figure}}

  Given a stream variable $u$ and position $i\geq 0$ an \emph{instant
    stream variable} (or simply instant variable) is defined as the
  pair $\InstVar{u}{i}$, which is a fresh variable of the same type as
  $u$.
  Note there is one different instant variable $\InstVar{u}{i}$ for
  each instant $i$.
  The \emph{evaluation graph} $\EvGraph$ is the unrolling expansion of
  the dependency graph for all instants.
  Given $\varphi(I,O)$ and a trace length $M$ (or $M=\omega$ for
  infinite traces) the evaluation graph $G_{\varphi,M}$ has as
  vertices the set of instant variables $\{\InstVar{u}{k}\}$ for
  $u\in I\cup O$ and $0\leq k<M$, and has edges
  $\InstVar{u}{k}\rightarrow \InstVar{v}{k'}$ if the dependency graph
  contains an edge $u\xrightarrow{j}v$ and $k+j=k'$

  The corresponding evaluation graph for $M = 5$ is shown in
  Fig.~\ref{plot:evalGraph}.

{\small\begin{figure}[tbh!]
  \centering
  \includegraphics[scale=0.3]{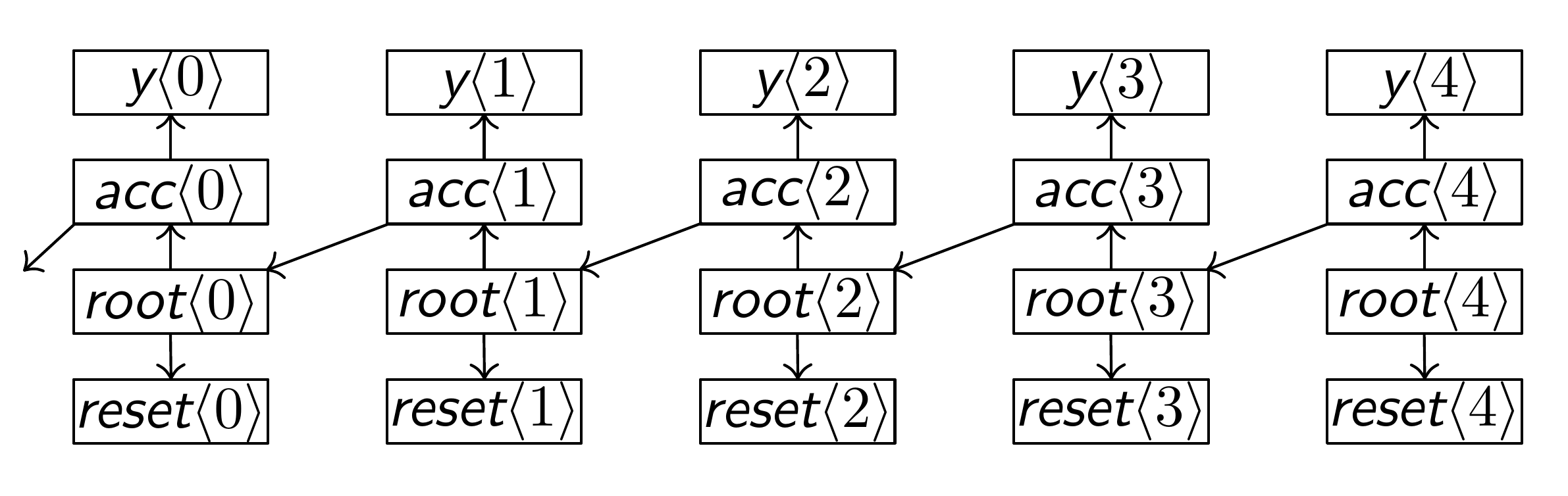} 
  \caption{Evaluation graph for example~\ref{ex:spec}}
  \label{plot:evalGraph}
  \end{figure}}

  % \vspace{0.5em}
  % {\small\hspace{2em} \includegraphics[scale=0.3]{evalGraph-crop} }
  
  Considering example~\ref{ex:spec}, $\InstVar{acc}{4}$ points to $\InstVar{root}{3}$
  in all evaluation graphs with $M\geq 4$.
  We denote by $\InstVar{e_s}{k}$ the term (whose leafs are instant
  variables) that results from $e_s$ at $k$, by replacing the offset
  terms with the corresponding instant variables corrected with the
  appropriated shift.
  Consider again Example~\ref{ex:spec}.
  The instant stream expression $e_{acc}$for $\StreamVar{acc}$ at
  instant $4$ is
  $\InstVar{acc}{4} = \InstVar{y}{4} + \InstVar{root}{3}$.

  Nodes of the dependency graph form a DAG of Maximal Strongly
  Connected Components (MSCCs).
  Note also that specifications whose dependency graph has no positive
  cycles are called \emph{efficiently monitorable
    specifications}~\cite{dangelo05lola}.
  There are no cycles in the evaluation graph of an efficiently
  monitorable specification, which enables us to reason by induction
  on evaluation graphs, as we will do later.
  Note that these specifications can have positive edges
  (corresponding to future dependencies) as long as they do not form a
  positive cycle.
  As it can be shown~\cite{sanchez18online} these specifications can
  be evaluated online (incrementally) with finite memory with a
  central monitor.
  \begin{example}
    The following code snippet shows a non-efficiently monitorable,
    an efficiently monitorable specification and a very efficiently monitorable.
    The first snippet is a non-efficiently monitorable specification
    because the stream b depends on itself in the future,
    in the Evaluation Graph (EG) all instant variables will
    depend on the next instant unboundedly to the future.
    This will make all instant streams b to never be resolved in an
    infinite trace.

    {\footnotesize\begin{lstlisting}[language=LOLA]
      input int a
      output int b = b[1|0]
    \end{lstlisting}}

  \noindent Next specification is an efficiently monitorable
  specification because there are only bounded references to the
  future: each instant variable b only depends on a two positions
  ahead, so for every instant variable $\InstVar{b}{k}$ it will be
  resolved at $k+2$.

    {\footnotesize\begin{lstlisting}[language=LOLA]
      input int a
      output int b = a[2|0] + b[-1|0]
    \end{lstlisting}}

  \noindent This is a very efficiently monitorable specification
  because there are no reference to the future, all offsets are either
  negative or zero.

    {\footnotesize\begin{lstlisting}[language=LOLA]
      input int a
      output int b = a + b[-1|0]
    \end{lstlisting}}

  \end{example}

\vspace{1em}  
  \subsection{Decentralized Synchronous Online Monitoring}
 %  \summary{explain dlola rv19 in short}
  An online decentralized algorithm to monitor \lola specifications in
  a synchronous networks is presented
  in~\cite{danielsson19decentralized}.
  The main idea is to use a network of cooperating nodes to monitor a
  \lola specification, sliced according to its syntax tree and then
  each subexpression, including inputs, is mapped to a node.
  This requires monitors to share their partial results (of the subexpressions)
  via messages.
  At each time instant the algorithm will read inputs, update internal
  expressions and communicate results with the appropiate nodes so
  that the specification ends being computed by means of those partial
  results.
  Therefore, given a well-formed \lola specification, the decentralized online
  algorithm presented in~\cite{danielsson19decentralized} incrementally
  computes the value for each output instant variable assuming a synchronous network
  where messages are not lost or duplicated.
  The algorithms presented here
  extend~\cite{danielsson19decentralized} to the more general setting
  of timed asynchronous networks.

\section{Decentralized Stream Runtime Verification for Timed Asynchronous Networks}
\label{sec:solution}
%\summary{overview of the solution: variables that we consider and the
%  big picture of the algorithm operation}

In this section we describe our solution to the decentralized SRV
problem for Timed Asynchronous Networks.
The algorithm that we present below will compute the unique values of
the output instant variables based on the values of the input
readings.
We prove the termination of the algorithm in theorem~\ref{thm:convergence}
and its correctness in theorem~\ref{thm:correctness}, verifying that the
operational semantics are equivalent to the denotational. 
We require a well-formed \lola specification, and a mapping between
streams and the network nodes where they are computed.
Each network node will host a local monitor that is responsible for
computing some of the streams of the
specification. 
We denote $\mu(s)$ for stream variable $s$ is the network node
whose local monitor is responsible for resolving the values of stream
$s$.
Local monitors exchange messages containing partial results whenever
needed in order to compute the global monitoring task.
However, our decentralized algorithm may compute some output values at
different time instants than a centralized version, due to the
different location of the inputs and the delays caused by the
communication.
We study this effect both theoretically in
Section~\ref{sec:resources-eager}, and empirically in
Section~\ref{sec:empirical}.
A centralized monitor corresponds with the operational semantics
in~\cite{dangelo05lola,sanchez18online} which is equivalent with a
network mapping that assigns all input and output streams to a single
node and therefore avoids communication.

%
% Structure:
% 3. Decentralized Stream Runtime Verification
%  Intro (above)
% 3.1 Problem Description
% 3.4 Dynamics (state and transition, both local and global)
% 3.5 Local Algorithm (Operational Semantics)
% 3.6 Correctness

\subsection{Problem Description}
\label{subsec:problem}
%\summary{inputs to our solution: network, spec and deployment}
% The description of the decentralized SRV monitoring problem consists
% of a specification, a network topology and a stream assignment.

\subsubsection{Network.}

We assume a network with a set of nodes $N$, such that every node can
communicate with every other network node by sending messages.
We assume reliable unicast communication (no message loss or
duplication) over a timed asynchronous network, so a given message can
take an arbitrary amount of time to arrive.
%
% In a degenerate example of this, consider a faulty link that for each
% time instant it presents a delay greater than the delay of the
% previous instant, and let's say it starts at time $0$ with a delay of
% $1$ time units.
% %
% Clearly, in this scenario we cannot find a global clock that allow us to
% build a synchronous network on top of this link, no matter how big we
% choose the clock tick to be.
% %
% This example shows that an asynchronous network is strictly more
% general than a synchronous because the synchronous is just a special
% case of asynchronous where the time varies by $0$ at each time
% instant.
%
Since network nodes share a global clock, the computation proceeds in
cycles.
In every cycle, all nodes in the network execute---in parallel and to
completion---the following actions: (1) read input messages, (2)
perform a terminating local computation, (3) generate output messages.
We use the following type of message: $(\InstVar{s}{k},c,n_s,n_d)$
where $\InstVar{s}{k}$ is an instant variable, $c$ is a value of the
type of $s$, $n_s$ is the source node and $n_d$ is the destination
node.
We use the following abbreviations $\MSG.\SRC=n_s$, $\MSG.\DST=n_d$,
$\MSG.\STREAM=\InstVar{s}{k}$ and $\MSG.\VAL=c$.
These messages are used to inform of the actual values read or
computed.

\subsubsection{Stream Assignment and Communication Strategy}
Given a specification $\varphi(I,O)$ and a network with nodes $N$, a
\emph{stream assignment} is a map $\mu:I\cup O\Into N$ that assigns a
network node to each stream variable.
The node $\mu(r)$ for an input stream variable $r$ is the location in
the network where $r$ is sensed in every clock tick.
At runtime, at every instant $k$ a new input value for
$\InstVar{r}{k}$ is read.
On the other hand, the node $\mu(s)$ for an output stream variable $s$
is the location whose local monitor is responsible for resolving the
values of $s$.

An instant value $\InstVar{v}{k}$ is automatically communicated to all
potentially interested nodes whenever the value of $\InstVar{v}{k}$ is
resolved.
Let $v$ and $u$ be two stream variables such that $v$ appears in the
equation of $u$ and let $n_v=\mu(v)$ and $n_u=\mu(u)$.
Then, $n_v$ informs $n_u$ of every value $\InstVar{v}{k}=c$ that $n_v$
resolves by sending a message $(\InstVar{v}{k},c,n_v,n_u)$.
We are finally ready to define the decentralized SRV problem.
\begin{definition}
  A decentralized SRV problem $\tupleof{\varphi,N,\mu}$ is
  characterized by a specification $\varphi$, a network with notes $N$
  and a stream assignment $\mu$ for every stream variable.
\end{definition}
We use DSRV for decentralized SRV problem.
Solving a DSRV instance consists of computing the values of instant
variables corresponding to the output streams based on the values of
the instant variables of the input streams, by means of a network of
interconnected nodes that host local monitors.

\subsection{Model of  Communication}
\label{subsec:time-async-model}
%\summary{timed asynch model of computation as we conceive it}
%
We now describe in detail the timed asynchronous model of computation
that we assume.
Every message inserted in the network arrive at its destination
according to the following conditions:
\begin{itemize}
\item \emph{Always later}: every message $m$ inserted at $t$ will
  arrive at $t'$ with $t' > t$;
\item \emph{Arbitrary delay}: there is no a-priori bound on the amount
  of time that any message will take to arrive.
\item \emph{FIFO between each pair of nodes}: let $m_1$ and $m_2$ be
  two messages with the same origin and destination,
  $m_1.\SRC=m_2.\SRC$ and $m_1.\DST=m_2.\DST$. Let $m_1$ is inserted
  at $t_1$ and arrive at $t_1'$ and let $m_2$ be inserted at $t_2$ and
  arrive at $t_2'$.  If $t_1<t_1$, then $t_1' \leq t_2'$.
  That is, $m_1$ cannot arrive later than $m_2$.
\end{itemize}
The synchronous model is a particular case of the timed asynchronous
in which all messages inserted in the network will always take the
same amount of time between each pair of network nodes.
In this case the delay will always be a constant.
Formally, to analyze the behavior of our algorithms we model the
message delays as a family of functions $\ArrivalTime{u}{v}$ (one for
each pair of nodes $(u,v)$, which provides at every moment $t$ the
instant $t'$ at which a message sent at $t$ from $u$ will arrive at
$v$.

\subsection{DSRV for Timed Asynchronous Networks: monitor and
  algorithm}
%\summary{monitor memory sets, informal description of algo}
Our solution consists of a collection of local monitors, one for each
network node $n$.
A local monitor $\tupleof{\QUEUE_n, U_n, R_n}$ for $n$ maintains an
input queue $\QUEUE_n$ and two storages:
\begin{compactitem}
\item \textbf{Resolved} storage $R_n$, where $n$ stores resolved
  instant variables $(\InstVar{v}{k},c)$.
\item \textbf{Unresolved} storage $U_n$, where $n$ stores unresolved
  equations $\InstVar{v}{k}=e$ where $e$ is not a value, but an expression that
  contains other instant variables.
\end{compactitem}

When $n$ receives a message from a remote node, the information is
added to $R_n$, so future local requests for the information can be
resolved locally and immediately.
At the beginning of the cycle of computation at instant $k$, node $n$
reads the values for input streams assigned to using local sensors and
instantiates for $k$ all output stream variables that $n$ is
responsible for.
After that, the equations obtained are simplified using the knowledge
acquired so far by $n$, which is stored in $R_n$.
Finally, new messages are generated and inserted in the queues of the
corresponding neighbors.

\AlgorithmTADLolaEager

More concretely, every node $n$ will execute the procedure \Monitor
shown in Algorithm~\ref{alg:alg-tadLola-eager}, which invokes \Step in
every clock tick.
The procedure \Finalize is used to resolve the pending values at the
end of the trace to their default.
Note that this procedure is never invoked if the monitor trace never
terminates (the monitor will be continuously observing and producing
outputs).
The procedure \Step executes the following steps:
\begin{enumerate}
\item \textbf{Process Messages}: Lines $7$ invokes \ProcessMessages
  procedure in lines $23$-$25$ that deals with the processing of
  incoming response arrivals, adding them to $R_n$
\item \textbf{Read Inputs and Instantiate Outputs:} Line $8$ reads new inputs
  for current time $k$, and line $9$ instantiates the equation of
  every output stream that $n$ is responsible for.
\item \textbf{Evaluate:} Line $10$ invokes the procedure \Evaluate, in lines $14-22$
  which evaluates the unresolved equations.
\item \textbf{Send Responses:} Line $12$ invokes \SendResponses, in lines $26$-$28$,
  sending messages for all newly resolved variables.
\item \textbf{Prune:} Line $29$-$31$ prunes the set $R$ from information
  that is no longer needed. See section~\ref{subsec:pruning}.
\end{enumerate}
%\clearpage
%
\subsection{Formal Correctness}
\label{subsec:correctness}
%\summary{formal proof of convergence and correctness}
We now show that our solution is correct by proving that the output
computed is the same as in the denotational semantics, and that every
output is eventually computed.

\newcounter{thm-convergence}
\setcounter{thm-convergence}{\value{theorem}}
\begin{theorem}
  \label{thm:convergence}
  All of the following hold for every instant variable $\InstVar{u}{k}$:
  \begin{enumerate}
    \item[\textup{(1)}] The value of $\InstVar{u}{k}$ is eventually resolved.
    \item[\textup{(2)}] The value of $\InstVar{u}{k}$ is $c$ if and only if $(\InstVar{u}{k},c)\in R$ at
      some instant.
    \item[\textup{(3)}] %If $\ISEAGER(u)$ then
      A response message for $\InstVar{u}{k}$ is eventually sent to
      all interested network nodes \textup{(}all nodes responsible for streams
      $v$ where $u\in\APPEARS(v)$\textup{)}.
  \end{enumerate}
\end{theorem}
\begin{proof}
  The proof proceeds by induction on the evaluation graph, showing
  simultaneously in the induction step $(1)$-$(3)$ as these depend on
  each other in the previous inductive steps.
  Let $M$ be a length of a computation (which can be $\omega$) and
  $\sigma_I$ be an input of length $M$.
  Note that $(1)$ to $(3)$ above are all statements about instant
  variables $\InstVar{u}{k}$, which are the nodes of the evaluation graph
  $G_{\varphi,M}$.
  We proceed by induction on $G_{\varphi,M}$ (which is acyclic because
  $D_\varphi$ is well-formed, by assumption).
  \begin{itemize}
  \item \textbf{Base case}: The base case are vertices of the
    evaluation graph that have no outgoing edges, which are either
    \begin{itemize}
    \item instant variables that correspond to inputs read from local
      sensors or
    \item to defined variables whose instant equation does not contain
      other instant variables;
      This is the case when either the equation is a constant or the
      time instant is such that the resulting offset falls off the
      trace; the default value is used.
    \end{itemize}
    Statement $(1)$ follows immediately for inputs because at instant
    $k$, $\InstVar{u}{k}$ is read at node $\mu(u)$.
    For output equations that do not have variables, or whose
    variables have offsets that once instantiated become negative or
    greater than $M$, the value of its leafs is determined either
    immediately or at $M$ when the offset is calculated.
    At this point, the value computed is inserted in $R$, so $(2)$
    also holds at $\mu(u)$.
    Note that $(2)$ also holds for other nodes because the response
    message contains $\InstVar{u}{k}=c$ if and only if
    $(\InstVar{u}{k},c)\in R_n$, where $\mu(u)=n$.
    %
    %If $s$ is eager,
    Then the response message is inserted exactly at
    the point it is resolved, so $(1)$ implies $(3)$.
    %
    %Finally, $(4)$ also holds at the time of receiving the request
    %message or resolving $\InstVar{u}{k}$ (whatever happens later).
%
  \item \textbf{Inductive case}: Consider an arbitrary
    $\InstVar{u}{k}$ in the evaluation graph $G_{\varphi,M}$ and let
    $\InstVar{u_1}{k_1},\ldots,\InstVar{u_l}{k_l}$ be the instant
      variables that $\InstVar{u}{k}$ depends on.
    These are nodes in $G_{\varphi,M}$ that are lower than $\InstVar{u}{k}$ so
    the inductive hypothesis applies, and $(1)$-$(3)$ hold for these.
    Let $n=\mu(u)$.
    At instant $k$, $\InstVar{u}{k}$ is instantiated and inserted in $U_n$. 
    %
    %At the end of cycle $k$, lazy variables among $u_1[l_1]\ldots
    %u_l[ul]$ are requested.
    %
    %By induction hypothesis, at some instant all these requests are
    %responded by $(1)$.
    %
    %Similarly, the values of all eager
    The values of instant variables are calculated and
    sent as well (by $(1)$ and $(3)$).
    At the latest time of arrival, the equation for $\InstVar{u}{k}$ has no more
    variables and it is evaluated to a value, so $(1)$ holds and $(2)$
    holds at $n$.
    At this point,
    %if $\ISEAGER(u)$ then
    the response message is sent (so $(1)$ holds for $\InstVar{u}{k}$) and %if $\ISLAZY(u)$ then all requests
    %(previously received in $\PEN_n$ or future requests) are answered,
    so $(1)$ also holds.
  \end{itemize}
  This finishes the proof.
\end{proof}

Theorem~\ref{thm:convergence} implies that every value of every
defined stream at every point is eventually resolved by our network of
cooperating monitors.
Therefore, given input streams $\sigma_I$, the algorithm computes (by $(2)$) the
unique output streams $\sigma_i$ one for each $s_i$.
The element $\sigma_i(k)$ is the value resolved for $\InstVar{s_i}{k}$
by the local monitor for $\mu(s_i)$.
The following theorem captures that
Algorithm~\ref{alg:alg-tadLola-eager} computes the right values
(according to the denotational semantics of Lola),
Theorem~\ref{thm:convergence} that all values are eventually
computed.% (the proof appears in the appendix).

We use $\OUTPUT(\sigma_I)$ as the function from input streams to
output streams that the cooperating monitors compute.
We use $\Osem{s}$ for the stream of values corresponding to stream
variable $s$ in $\OUTPUT(\sigma_I)$.
We now show that the sequence of values computed corresponds to the
semantics of the specification.
\newcounter{thm-correctness}
\setcounter{thm-correctness}{\value{theorem}}
\begin{theorem}
  \label{thm:correctness}
  Let $\varphi$ be a specification, $S=\tupleof{\varphi,\topo,\mu}$ be
  a decentralized SRV problem, and $\sigma_I$ an input stream of
  values.
  Then $(\sigma_I,\OUTPUT(\sigma_I))\models \varphi$.
\end{theorem}

% PROOF in appendix
\begin{proof}
  Let $\sigma_O$ be the unique evaluation model such that
  $(\sigma_I,\sigma_O)\models \varphi$ (we use $\sigma_O(s)$ for the
  output stream for stream variable $s$ and $\sigma_O(s)(k)$ for its
  value in the $k$-th position).  
  We need to show that for every $s$ and $k$,
  $\Osem{s}(k)=\sigma_O(s)(k)$.
  We again proceed by induction on the evaluation graph $G_{\varphi,M}$.
  \begin{itemize}
  \item \textbf{Base case:}
    For inputs the value follows immediately. 
    The other basic case corresponds to output variables $s$ at
    instants at which these that do not depend on other variables
    (because all occurrences of offsets, if any, fall off the trace).
    The evaluation of the value is performed by network node $\mu(s)$, and it
    satisfies the equation $e_s$ of $s$, not depending on any value of
    any other stream. 
    Therefore, it satisfies that $\Osem{s}(k)=\sem{e_s[k]}=\sigma_O(s)(k)$,
    as desired
  \item \textbf{Inductive case:} 
    Let $s$ be an arbitrary stream variable and $k$ an arbitrary
    instant within $0$ and $M-1$ and assume that all instant variables
    $\InstVar{u}{k'}$ that $\InstVar{s}{k}$ can reach in the evaluation graph satisfy the
    inductive hypothesis.
    Let $n$ be the node in charge of computing $s$.
    By Theorem~\ref{thm:convergence}, all the values are eventually
    received by $n$ and in $R_n$, and by IH, these values are the same
    as in the denotational semantics, that is
    $\Osem{u}(k')=\sigma_O(u)(k')$.
    The evaluation of $\InstVar{s}{k}$ corresponds to computing $\sem{e_s}$,
    which uses the semantics of the expression (according to
    Section~\ref{sec:lola}).
    A simple structural induction on the expression $e_s$ shows that
    the result of the evaluation, that is the value assigned to
    $\InstVar{s}{k}$, is $\sem{e_s}_{\sigma}(k)=\sigma_O(s)(k)$, as desired.
  \end{itemize}
  This finishes the proof.
\end{proof}
\subsection{Simplifiers}
\label{sc:simplification}
%\summary{simplifiers, just as in DSRV}
%
The evaluation of expressions in Algorithm~\ref{alg:alg-tadLola-eager}
assumes that all instant variables in an expression $e$ are known (\ie
$e$ is ground), so the interpreted functions in the data theory can
evaluate $e$.
Sometimes, expressions can be partially evaluated (or even the value
fully determined) knowing only some but not all of the instant variables
involved in the expression.
As simplifier is a function $f: \Term_D \Into \Term_D$ such that (1)
the variables in $f(t)$ are a subset of the variables in $t)$, and
(2) every substitution of values for the variables of $t$ produces the
same value as the substitution of $f(t)$.
For example, the following are typical simplifiers:
\[
  \begin{array}{lcl@{\hspace{3em}}lcl@{\hspace{3em}}lcl}
    \textit{if } \;\;\true\;\; \textit{ then } \;\;t_1\;\; \textit{ else } \;\;t_2 & \mapsto & t_1 \\
    \textit{if } \;\;\false\;\; \textit{ then } \;\;t_1\;\; \textit{ else } \;\;t_2 & \mapsto & t_2 \\
    \;\true\; \Or x & \mapsto & \true \\
    \;\true\; \And x & \mapsto & x \\
    0 \cdot x & \mapsto & 0 \\
  \end{array}  
\]
In practice, simplifiers can dramatically affect the performance in
terms of the instant at which an instant variable is resolved and, in
the case of decentralized monitoring, the delays and number of
messages exchanged.
Essentially, a simplifier is a function from terms to terms such that,
for every possible valuation of the variables in the original term it
does not change the final value obtained.
%
% The following follows immediately.
% %
% \begin{fact}
%   Let $\varphi(I,O)$ be a well-formed specification, $\sigma_I$ be an
%   input valuation and $\sigma_O$ the corresponding output valuation.
%   %
%   Then for every term $e\in\Term_D(I\cup O)$ and every simplifier $f$,
%   $\sem{e}_{\sigma_I,\sigma_O}=\sem{f(e)}_{\sigma_I,\sigma_O}$.
%   \label{fact:simplifier-preserves}
% \end{fact}
%
It is easy to see that for every term $t$ obtained by instantiating a
defining equation and for every simplifier $f$,
$\sem{t}_{\sigma_I,\sigma_O}=\sem{f(t)}_{(\sigma_I,\sigma_O)}$,
because the values of the variables in $t$ and in $f(t)$ are filled
with the same values (taken from $\sigma_I$ and $\sigma_O$).
%
% The following also holds for every $\varphi$ and valuation
% $(\sigma_I,\sigma_O)$.

% \begin{lemma}
%   Let $e$ be an instant term and let $\rho=\{\InstVar{u}{k}\mapsto c\}$ be the
%   substitution such that $c=\sem{\InstVar{u}{k}}_{(\sigma_I,\sigma_O)}$.
%   %
%   Then, $\sem{e}_{(\sigma_I,\sigma_O)}=\sem{e\triangleleft \rho}_{(\sigma_I,\sigma_O)}$.
%   \label{lem:application-preserves}
% \end{lemma}
% %
% Lemma~\ref{lem:application-preserves} holds immediately because the
% substitution $\rho$ is just the partial application of one of the
% values of the variables that may appear in $e$.
%
Consider arbitrary simplifiers $\simp$ used in line $19$ of
Algorithm~\ref{alg:alg-tadLola-eager} to simplify expressions.
Let $U_n$ be the unresolved storage for node $n$ and let $\InstVar{u}{k}$ be an
instant variable with $\mu(u)=n$.
By Algorithm~\ref{alg:alg-tadLola-eager} the sequence of terms
$(\InstVar{u}{k},t_0), (\InstVar{u}{k},t_1), \ldots
(\InstVar{u}{k},t_k)$ that $U_n$ will store are such that each $t_i$
will have the simplifier applied.
It follows that the value computed using simplifiers is the same as
without simplifiers.
It is also easy to show that the algorithm using simplifiers obtains
the value of every instant variable no later than the algorithm that
uses no simplifier.
This is because in the worst case every instant variable is resolved
when all the instant variables it depends on are known, and all
response messages are sent at the moment they are resolved.

\subsection{Theoretical Resource Utilization}
\label{sec:resources-eager}
%\luismitext{Add plots of network behavior in old Journal dir}
%\summary{recall decentralized efficiently monitorable}
The aim of this section is to define conditions under which local
monitors only need bounded memory to compute every output value.
The first thing to consider is that the specification must be
\emph{decentralized efficiently
  monitorable}~\cite{danielsson19decentralized}, which essentially
states that every strongly connected component in $G_\varphi$ must be
mapped to the same network node.
That is, if $u$ appears, transitively, in the declaration of $v$ and
$v$ appears in the declaration of $u$ (with some offsets), then
$\mu(u)=\mu(v)$.

In order to guarantee that a given storage in a local monitor for node
$n$ is bounded, we must provide an upper-bound for how long it takes
to resolve an instant variable for a stream that is assigned to $n$.
We use Time to Resolve (TTR) to refer to the ammount of time that
a given instant variable $\InstVar{u}{k}$ takes to get resolved.
This is the number of time instants between the instantiation of the
variable at time $k$ and the instant at which it gets resolved,
leaving $U_n$ and being stored in $R_n$.
This happens in line $21$ in Algorithm~\ref{alg:alg-tadLola-eager}.

\subsubsection{General Equations for the Time to Resolve}
We introduce now a general definition of recursive equations that
capture when an instant variable $\InstVar{s}{k}$ is resolved.
In order to bound the memory used by the monitor at network node $n$,
we need to bound storages $U_n$ and $R_n$:
%, since all the other storages depend on the them:
%
\begin{itemize}
\item \textbf{Bound on $R_n$}: %Under the eager strategy
  Resolved values that are needed remotely are sent immediately to the
  remote nodes, so $R_n$ only contains resolved values that are needed
  in the future locally at $n$.
  Since efficiently monitorable specifications only contain (future)
  bounded paths there is a maximum future reference $b$ used in the
  specification.
  This upper-bound limits for how long a resolved value
  $\InstVar{v}{k}$ can remain in $R_n$, because after at most $b$
  steps the instant variables $\InstVar{u}{k'}$ that need the value of
  $\InstVar{v}{k}$ stored in $R_n$ will be instantiated (note that
  $k'-k\leq b$).
  %
  % \luismitext{In fact what happens in the implementation is that the
  %   value is sent, the dst monitor stores it and uses it once the
  %   value gets instanced}

  That is $\InstVar{u}{k}$ is not needed after
  $t=max(k+b, k+TTR(\InstVar{u}{k}))$.
  At $t$, the value of $\InstVar{u}{k}$ can be removed from $R_n$.
  This guarantees that the size of $R_n$ is always upped-bounded by a
  constant in every node $n$.
\item \textbf{Bound on $U_n$}: The size of the memory required for
  storage $U_n$ at the node $n$ responsible to resolve $s$ (that is
  $n=\mu(s)$) is proportional to the number of instantiated but
  unresolved instant variables.
  Therefore, to bound $U_n$ we need to compute the bound on the time
  it takes to resolve instant variables of streams assigned to $n$.
\end{itemize}
The general equations that we present below depend on the delay of
messages in the network.
We will later instantiate these general equations for the following
cases of network behavior:
\begin{itemize}
\item a synchronous network;
\item a timed-asynchronous network with an upper-bound on message
  delays for the whole trace (we call this the \globalTime case);
\item timed-asynchronous network with an upper-bound for message
  delays in a given time-horizon (we call this the \localTime case).
\end{itemize}
Note that the correctness of the algorithm
(Theorem~\ref{thm:correctness}) establishes that the output streams
$\sigma_O$ only depend on the input streams $\sigma_I$ but does not
state bounds on the time at which each element of $\sigma_O$ is
resolved or on the delays of messages.

In this section we study how the delay of messages affects the time at
which instant variables are resolved, which in turn affects the memory
usage at the computations nodes.
We use $\TAmsgTrip(t,a,b)$ for the time it takes for a message sent
from $a$ to $b$ at time $t$ to arrive.
In other words $\ArrivalTime{a}{b}(t)= t + \TAmsgTrip(t,a,b)$.
Recall that we assume that messages are causal and queues are FIFO as
we described in~\ref{subsec:time-async-model}.
Causality means that messages arrive after they are sent (that is, for
every $n$, $m$ and $t$, $\ArrivalTime{a}{b}(t) > t$) and FIFO that for
every $n$ and $m$, if $t<t'$ then
$\ArrivalTime{a}{b}(t) \leq \ArrivalTime{a}{b}(t')$.

We now capture the \emph{Moment to Resolve} for a given instant
variable $\InstVar{s}{t}$, represented as $\MTR(\InstVar{s}{t})$,
which captures the instant of time at which $\InstVar{s}{t}$ is
guaranteed to be resolved by the monitor at network node $\mu(n)$
responsible to compute $s$.
Our definition considers two components, the delay in resolving all
local instant variables that $\InstVar{s}{t}$ may depend on and the
resolution of remote instant variables, which also involve message
delays.
We use the concept of \emph{remote moment to resolve}, denoted
$\MTRToRem{\MTR}(\InstVar{s}{t})$, as the instant at which all remote
values that $\InstVar{s}{t}$ directly require have arrived (which is
$t$ if all values arrive before $t$).
\begin{align*}
& \MTR(\InstVar{s}{t}) \DefinedAs \max( \\
& \hspace{1em}
    \MTRToRem{\MTR}(\InstVar{s}{t}),\{\MTR(\InstVar{r}{t+w})\;|\; \depWeightSub{s}{w}{loc}{r}\}) \\  
& \MTRToRem{\MTR}(\InstVar{s}{t}) \DefinedAs \max(t,\\
& \hspace{1em}\{ \ArrivalTime{r}{s}(\MTR(\InstVar{r}{t+w})) \;|\;\depWeightSub{s}{w}{rem}{r} \text{ and } t+w \geq 0 \}) \\
\end{align*}
Note that this is well-defined for every well-formed specification
because the evaluation graph is acyclic, and the equation for
$\InstVar{s}{t}$ only depends on those variables lower in the
evaluation graph, which is acyclic.

\begin{example}
  Consider example~\ref{ex:spec} with streams $i$ and
  $acc$ at network node $1$ and streams $\RESET$ and $\ROOT$ computed at
  network node $2$.  %
  Then, we can substitute in the equations to obtain the
  $\MTR(\InstVar{\ROOT}{1})$.
\begin{align*}
  &  \MTR(\InstVar{\ROOT}{1}) = \max(\MTR(\InstVar{\RESET}{1}), \MTRToRem{\MTR}(\InstVar{\ACC}{1})) = \\
  & = \max(1, \max(1,\ArrivalTime{\ACC}{\ROOT}(\MTR(\InstVar{\ACC}{1})))) = \\ 
  & = \max(1, \ArrivalTime{\ACC}{\ROOT}(\max(1,\\
  & \hspace{3em}\MTR(\InstVar{i}{1}), \MTRToRem{\MTR}(\InstVar{\ROOT}{0})))) =\\ 
  & = \max(1, \ArrivalTime{\ACC}{\ROOT}(\max(1, \MTRToRem{\MTR}(\InstVar{\ROOT}{0}))))= \\
  & = \max(1, \ArrivalTime{\ACC}{\ROOT}(\max(1,\\
  & \hspace{3em}\max(\MTR(\InstVar{\RESET}{0})\MTRToRem{\MTR}(\InstVar{\ACC}{0}))))) = \\
  & = \max(1, \ArrivalTime{\ACC}{\ROOT}(\max(1, \max(0, \\
  & \hspace{3em}\max(0,\ArrivalTime{\ACC}{\ROOT}(\MTR(\InstVar{\ACC}{0}))))))) = \\
  & = \max(1, \ArrivalTime{\ACC}{\ROOT}(\max(1, \ArrivalTime{\ACC}{\ROOT}(\max(0,\\
  & \hspace{3em}\MTR(\InstVar{i}{0}), \MTRToRem{\MTR}(\InstVar{\ROOT}{-1})))))) = \\
  & = \max(1, \ArrivalTime{\ACC}{\ROOT}(\max(1, \ArrivalTime{\ACC}{\ROOT}(0))))\\
\end{align*}

The instant variable $\InstVar{\ROOT}{1}$ is guaranteed to be resolved
when the response from the instant variable $\InstVar{\ACC}{1}$
arrives---that is the $\max(1, \ArrivalTime{\ACC}{\ROOT}(...))$ part.
And this response can only be produced when the response for
$\InstVar{\ACC}{0}$ is arrives, which is the innermost part:
$...\max(1, \ArrivalTime{\ACC}{\ROOT}(0))$
Note that we do not need to account for
$\MTRToRem{\MTR}(\InstVar{root}{-1})$ since it is resolved
instantaneously to its default value.  Likewise, the inputs are also
resolved instantaneously and do not add any delay when obtaining the
value of the $\MTR$.
\end{example}

Note that for $\MTRToRem{\MTR}(\InstVar{s}{t})$ only consider the
those remote instant variables for which $t+w\geq{}0$ because
otherwise the default value will be used at the moment of
instantiating $\InstVar{s}{t}$.
In the equation for $\MTR(\InstVar{s}{t})$ we assume the base case
$\MTR(\InstVar{s}{t})=0$ when $t<0$, because again, the default value
in the offset expression is used instead, which is known immediately.
It is easy to see that the first equation is equivalent to:
\[  \MTR(\InstVar{s}{t}) \DefinedAs 
    \max(\{\MTRToRem{\MTR}(\InstVar{r}{t+w})\;|\; \depWeightSubStar{s}{w}{loc}{r}\}) 
\]
We are now ready to prove that these definitions indeed capture the time at
which $\InstVar{s}{t}$ is resolved.

\begin{theorem}
  Let $\Spec$ be a specification and $\mu$ a network
  placement, let $\sigma_I$ be the input trace and $\Arr$
  a network behavior.
  Every $\InstVar{s}{t}$ is resolved at $\MTR(\InstVar{s}{t})$ or
  before.
\end{theorem}

\begin{proof}
  The proof proceeds by induction on the evaluation graph
  $G_{\varphi,M}$ induced by $\Spec$ and the length of $\sigma_I$.
  \begin{itemize}
  \item \textbf{Base case}: inputs and instant variables
    $\InstVar{s}{t}$ that do not depend on any other instant
    variables.
    These are the nodes of $\EvGraph$ that do not have any outgoing
    edge.
    Since $\InstVar{s}{t}$ is instantiated at $t$, then the
    value is resolved exactly at $t$ either by reading a sensor or
    instancing to a default value.
    Also, $\MTR(\InstVar{s}{t})=\MTRToRem{\MTR}(\InstVar{s}{t})=t$.
  \item \textbf{General case}. Let $\InstVar{s}{t}$ be an arbitrary
    instant variable and assume, by inductive hypothesis, that the
    theorem holds for all instant variables lower in the $\EvGraph$
    than $\InstVar{s}{t}$.
    At time $\MTRToRem{\MTR}(\InstVar{s}{t})$ all instant variables
    $\InstVar{r}{t+w}$ from remote nodes that $\InstVar{s}{t}$ depends
    on have arrived because $\InstVar{r}{t+w}$ will be resolved at
    $\MTR(\InstVar{r}{t})$ by induction hypothesis.
    Similarly, all local elements that $\InstVar{s}{t}$ depends on are
    also below in the dependency graph, so the induction hypothesis
    also applies.
    Therefore, at time
    \[ \max(\MTRToRem{\MTR}(\InstVar{s}{t}),\{\MTR(\InstVar{r}{t+w})\;|\;\depWeightSub{s}{w}{loc}{r}\}\} \]
    or before all elements that $\InstVar{s}{t}$ depends on will be
    known and $\InstVar{s}{t}$ will be resolved.
  \end{itemize}
  This finishes the proof.
\end{proof}

The following corollary follows from the fact that nothing that
happens after an instant variable has been resolved (either further
values in $\sigma_I$ or the network behavior) can affect the value
computed.
Therefore, the value and time at which $\InstVar{s}{t}$ is computed
does not depend on the future after $\MTR(\InstVar{s}{t})$.

\begin{corollary}
  For all $\InstVar{s}{t}$ there is a $t'$ such that $\InstVar{s}{t}$
  only depends on $\sigma_I$ and $\Arr$ up to $t'$.
\end{corollary}

The $\MTR$ for an instant variable depends on the delay of the network
$\ArrivalTime{}{}$ between the network nodes that cooperate in order
to compute that instant variable.
Therefore we cannot guarantee a bound on $\MTR$ if those delays can be
arbitrarily long, so we cannot bound the memory usage.
Consequently, monitoring is not trace-length independent in a general
Time Asynchronous Network.

Next, we study how different conditions on the network behavior
(concerning the delays between links) affect the $\MTR$ establishing
memory bounds and regain trace-length independent monitoring under
those conditions.

\subsubsection{Instantiation to Synchronous Time}
% \luismitext{Synch: calculation of TimeToResolve == Look-ahead}
% \luismitext{In centralized synch(LOLA) the concept of TimeToResolve
% collides to Latency which is based on the offsets in the Depgraph}

We assume first the synchronous model of computation, which is a
particular case of the timed-asynchronous model where all message
delays between two monitors take exactly the same amount of time
throughout the trace.
%
% And that this constant delays are known a priori.
%
We use the predicate $\DIST{r}{s}$ to represent the delay that every
message will take from $\mu(r)$ to $\mu(s)$, independently of the time
instant at which the message is sent.
Therefore $\ArrivalTime{r}{s}(t) = t + \DIST{r}{s}$.
This delay allows us to simplify $\MTRToRem{\MTR}$ for synchronous
networks as follows:
\begin{align*}
  &\MTRToRem{\MTRsyn}(\InstVar{s}{t}) \DefinedAs \max(t,     M(\InstVar{s}{t}))\\
  \intertext{where $M(\InstVar{s}{t})=$}
  &\{ \MTRsyn(\InstVar{r}{t+w}) + \DIST{r}{s} \;|\;\depWeightSub{s}{w}{rem}{r}, t+w \geq 0 \})
\end{align*}
Recall that the \emph{time to resolve} is the time interval between
the moment at which a variable is instantiated and the instant at
which it is resolved (that is
$\TTR(\InstVar{s}{t}) = \MTR(\InstVar{s}{t}) - t)$
In the synchronous case we obtain:
\begin{align*}
  & \TTRsyn(\InstVar{s}{t}) = \MTRsyn(\InstVar{s}{t}) - t = \\
  &= \{\MTRToRem{\MTRsyn}(\InstVar{r}{t+w}) \;|\; \depWeightSub{s}{w}{loc}{r}\} -t = \\
  % &= \max(t, \{ \MTRsyn(\InstVar{r}{t+w}) + \DIST{r}{s} \;|\; \depWeightSub{s}{w}{rem}{r} \}) - t =\\
  &= \max(t, M(\InstVar{s}{t}) - t = \\
  &= \max(0,\{\TTRsyn(\InstVar{r}{t+w}) + \DIST{r}{s} \;|\; \depWeightSub{s}{w}{rem}{r}\})
\end{align*}
Note that the value that determines the result is the
$\TTRsyn(\InstVar{s}{t})$ of the slowest remote dependency, which
includes both the resolve time and the time the message needs to
traverse through the network.
Additionally, we can easily show by induction on the dependency graph
that for every stream variable $s$ there is a constant $k$ such that
$\TTRsyn(\InstVar{s}{t})\leq k$, that is, $s$ always takes less than
$k$ instants to be resolved.
%
% This is equivalent to the result obtained
% in~\cite{danielsson19decentralized}.
%
It follows that all decentralized efficiently monitorable
specifications can be monitored in constant space in every local
monitor, that is, synchronous decentralized monitoring of
decentralized efficient monitorable specifications is trace length
independent.
%
% \luismitext{Corollary?}
% This comfirms our assumption that the synchronous network behavior
% is a particular case of an asynchronous network behavior.

\subsubsection{Timed Asynchronous with \GlobalTimeAdv Bounded delays}
We now assume that there is a global upper bound on the delay time for
every message, which we call \globalTimeAdv bounded delays.
%
%% NOT REFERRED
% {\small\begin{figure}[tbh!]
%   \centering
%   \includegraphics[scale=0.7]{plots/ttr_8_growing_try-crop.pdf} 
%   \caption{Network behavior that is not globally bounded}
%   \label{plot:growingTTA}
% \end{figure}}
%
Formally, this assumption states that if there is a $d$ such that for
every pair of streams $r,s$ and for every time $t$,
$\ArrivalTime{r}{s}(t)\leq t+d$.
Substituting the upper-bound value $d$ in the equations for $\MTR$, we
obtain an constant upper-bound on the $\MTR$:
\begin{align*}
  &\MTRToRem{\MTRg}(\InstVar{s}{t}) \leq \max(t,
    M(\InstVar{s}{t}))\\
  \intertext{where}
  & M(\InstVar{s}{t})=\MTRg(\InstVar{r}{t+w}) + d \;|\;\depWeightSub{s}{w}{rem}{r}, t+w \geq 0 \}
\end{align*}
Note that in some cases $\InstVar{s}{t}$ can be resolved before
$\MTRg(\InstVar{s}{t})$ because $d$ is an upper bound.
In this case we can also bound the memory necessary to store in every
node to perform the monitoring process, but most of the time less
memory will be necessary.
We can see an example of a \globalTime bound in
Figure~\ref{plot:peakTTR}.

{\small\begin{figure}[tbh!]
  \centering
  \includegraphics[scale=0.7]{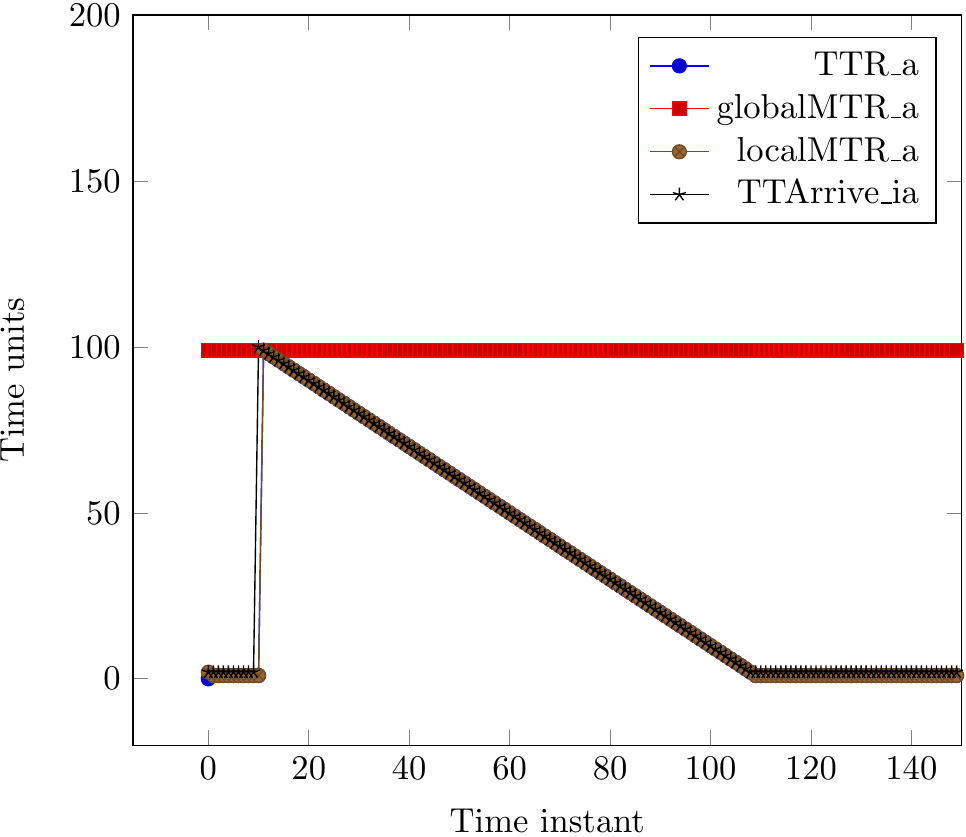} 
  \caption{TTR, \globalTime and \localTime bounds for a peak delay}
  \label{plot:peakTTR}
\end{figure}}

\subsubsection{Timed Asynchronous with \LocalTimeAdv Bounded delays}
%
%\cesartext{is this an assumption or the more general case?}
%\luismitext{it is the most general case because it works even if the network delay is as bad as: for every t, arr(t+1) > arr(t), there is always a local bound!}
%
We now take a closer look at the equations to obtain a better bound on
the time to resolve a given instant variable $\InstVar{s}{t}$, without
assuming an upper-bound of all messages in the history of the
computation, but only the necessary messages that can influence
$\InstVar{s}{t}$.
The main idea to bound $\MTR(\InstVar{s}{t})$ is to consider the time
interval at which the messages that are relevant to compute
$\InstVar{s}{t}$ are sent.
We first define an auxiliary notion.
We say that a stream variable $r$ is a direct remote influence on $s$
with delay $w$, and we write $\wdirrem{s}{w}{r}$, whenever there is a
path
$\wloc{s}{w_1}{s_1}\wlocnext{w_2}{s_2}\ldots\wlocnext{w_k}{s_k}\wremnext{w_{k+1}}r$
such that:
\begin{compactitem}
\item no two nodes $s_i$ and $s_j$ are repeated (if $i\neq j$ then $s_i\neq s_j$), and
\item $w=w_1+\ldots+w_k+w_{k+1}$.
\end{compactitem}
Note that $\wdirrem{s}{w}{r}$ means that $\InstVar{s}{t}$ may be
influenced by remote variable $\InstVar{r}{t+w}$.
We define the window of interest for $\InstVar{s}{t}$ as:
\begin{align*}
  &\Win(\InstVar{s}{t}) = [\min S, \max S] \text{ where $S$ is defined as } \\
  &S  =  \{t, \MTRToRem{\MTR}(\InstVar{r}{t+w}) \;|\; \wdirrem{s}{w}{r} \text{ and } t+w>0\}
\end{align*}
Note that $S$ is the set of instants at which remote instant variables
that influence $\InstVar{s}{t}$ are sent.
\begin{example}

  Considering the specification in example~\ref{ex:spec}
  and by taking a look at the evaluation graph in
  Figure~\ref{plot:evalGraph} we observe that the window of interest of
  the any instant variable at any time includes those of its
  dependencies in the evaluation graph.
  Therefore, their window of interest will include
  the minimum time for the earliest dependency
  to be resolved and the maximum time for the last dependency to be
  resolved.
  In this example, the window for $\InstVar{root}{1}$ will include the windows for
  $\InstVar{acc}{1}$, $\InstVar{root}{0}$ and $\InstVar{acc}{0}$ and the time required for the
  response messages to travel from source to destination.
  Note that inputs do not affect the $\MTR$.
  \end{example}
Therefore $\Win(\InstVar{s}{t})$ contains those instants at which the
remote information relevant to $\InstVar{s}{t}$ is sent.
This window always ends at most at $\MTR(\InstVar{s}{t})$.
We then define the worst message sent to $s$ for the computation of
$\InstVar{s}{t}$ as:
\begin{align*}
  &\dworst(\InstVar{s}{t})=\max\{t'-t\;| \\
  &\hspace{3em}t'=\ArrivalTime{r}{s}(t) \text{ for }  \wdirrem{s}{w}{r} \text{ and  } t\in \Win(\InstVar{s}{t}) \}.
\end{align*}
Note that $\dworst$ is still an over-approximation of the messages
sent in order to compute $\InstVar{s}{t}$ but in this case the bound considers all those
messages and only looks at a bounded interval of time.
Since all the values that influence $\InstVar{s}{t}$ are sent within
$\Win(\InstVar{s}{t})$ we can bound $\MTR(\InstVar{s}{t})$ as follows:
\begin{align*}
  &\MTRToRem{\MTRlocal}(\InstVar{s}{t}) \leq \max(t,M(\InstVar{s}{t}))\\
  \intertext{where $M(\InstVar{s}{t})=$}
  &\{\MTR(\InstVar{r}{t+w})+\dworst(\Win(\InstVar{s}{t}))\;|\;\wdirrem{s}{w}{r} \}.
\end{align*}
We have finally arrived at the desired outcome: a finite window of
time that contains the sending and receiving of the relevant messages
for the computation of the instant variable.
This implies that only a finite number of network delays affect the
resolution of any instant variable $\InstVar{s}{t}$.
As we can always find the maximum delay in the window, we can upper
bound the time that it will take for any instant variable to be
resolved, and we are able to know how much time these instant
variables are stored in $U_n$ and $R_n$.
In turn, this allows to determine when certain instant variables are
no longer needed and when they can be pruned releasing the used
memory.

Figure~\ref{plot:peakTTR} shows the peak network behavior and how the
TTR adapts accordingly. We can observe the difference between the
\localTime and \globalTime bounds, where the \globalTime bound is high
and constant throughout the execution and the \localTime drops when
the network has small delays.

\subsubsection{Pruning the Resolved Storage $R_n$.}
\label{subsec:pruning}
% \summary{see RV19 part where we talk about pruning and comment
% pruning algo~\ref{alg:prune}}
%
We are finally ready to prune $R_n$ because we know now when every
instant variable will be resolved.
%
% \luismitext{DANGER!!!: as we are considering eager only this whole
% discussion about how to prune is pointless(this whole section!!!)
% since with just a fire-n-forget strategy it works}

\begin{corollary}
  \label{thm:mtr}
  Every unresolved instant variable $\InstVar{s}{t}$ in $U_n$ is resolved at most at
  $\MTR(\InstVar{s}{t})$.
\end{corollary}
As soon as $\MTR(\InstVar{s}{t}$ is reached (or before), the value of
$\InstVar{s}{t}$ will be known in the local monitor of $\mu(s)$ and
its value will be sent to those remote monitors where it is needed.
After this moment $\InstVar{s}{t}$ can be pruned from $U_n$.
With this mechanism, we can assure that every instant variable will be
in memory ($U_n$ or $R_n$) for a bounded amount of time.
Corollary~\ref{thm:mtr} implies that decentralized efficiently
monitorable specifications in timed asynchronous networks can be
monitored with bounded resources when there is a certain bound on the
network behavior, be it synchronous, \globalTime or a \localTime
bound.
This memory bound depends only linearly on the size of the
specification and the delays between the nodes of the network.
This results can be interpreted from the opposite perspective: given a
fixed amount of memory available, we could calculate the maximum
delays in the network that would allow the monitoring to be performed
correctly.

%%% Local Variables:
%%% TeX-master: "main.tex"
%%% TeX-PDF-mode: t
%%% End:

%\input{efficient}
%\input{examples}
\section{Empirical Evaluation}
\label{sec:empirical}
%\summary{intro to empirical: implementation and hypothesis}
%\TableResults
%\TableResultsTadLola
%
\TableTADlolaTTRAnalysis
We have implemented our solution in a prototype tool tadLola, written
in the Go programming language (available at
\url{http://github.com/imdea-software/dLola}).
We describe now:
\begin{compactitem}
\item (1) an empirical study of the capabilities of tadLola in
  different scenarios with real data extracted from four different
  realistic public datasets.
\item (2) the effect of the network behavior---in terms of
  delays--into memory and time to resolve outputs.
\end{compactitem}
Our experimental setup intends to empirically determine the behavior
of the asynchronous network and how failures affect the time to resolve
of the streams.

\subsection{Datasets and Network Failures.}
We have used four different datasets for this empirical evaluation,
namely: SmartPolitech~\cite{pajuelo-holguera2020smartPolitech}, Tomsk
Heating~\cite{zorin2020tomskHeating},
Orange4Home~\cite{cumin17orange4home} and
Context~\cite{kaupp2021contextDataset}.
All datasets are related to smart buildings except for Context that is about Industry 4.0.
The first two are concerned about building climate control and use
sensors in different rooms or buildings respectively.
Orange4Home dataset focus on activity recognition where a tenant
can move freely in an apartment, and the goal is to infer the activity
performed.
Lastly, Context is a dataset in a smart factory where a new class of failures,
namely contextual failures arise when there is no specific sensor or data collected that signals directly the error
but the presence of the error and its underlying cause need to be inferred from contextual knowledge.
For each dataset we created a synthetic specification that could
showcase the functionality of our tool.
We also injected synthetic delays to model network congestion and
failures.

\begin{itemize}
\item \textit{constant} behavior is modeled as a global constant delay between
  each pair of monitors, so every message takes exactly a fixed amount of
  time to reach the destination network node.
  This corresponds to the network behavior observable in synchronous
  monitoring.
\item \textit{constantPeak} consists of a constant delay with a single
  high delay of the network modeling a network failure and recovery,
  so all messages get delayed until the problem is solved and then the
  network starts to recover gradually, until normal operation is
  reached again.
\item \textit{Normal} behavior follows a normal distribution of the
  delays given an average delay.
\item \textit{normalPeak} is similar to the constantPeak but with a
  baseline of the normal behavior.
\end{itemize}

Note that all these behaviors are both \globalTime and \localTime bounded
since for all of them we can find an upper bound for the whole trace as well as
a bound by window of interest of each instant variable.

Figure~\ref{fig:tadLola-ttr-analysis} shows the minimum, median and
maximum TTR to resolve streams under these network behaviors.

\begin{figure}[tbh!]
  \begin{tabular}{l}
  \includegraphics[scale=0.6]{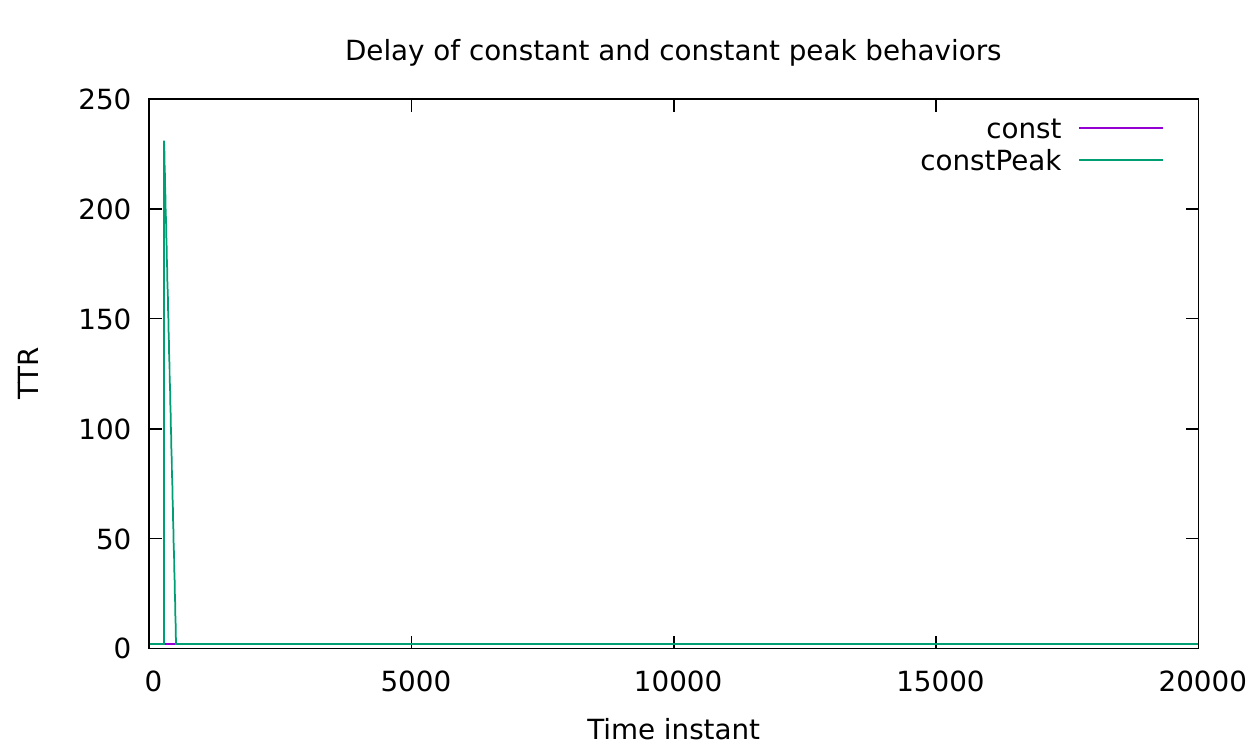} \\
  \multicolumn{1}{l}{\begin{tabular}{l}\\[-0.5em](a) Const and const Peak \end{tabular}}\\
  \includegraphics[scale=0.6]{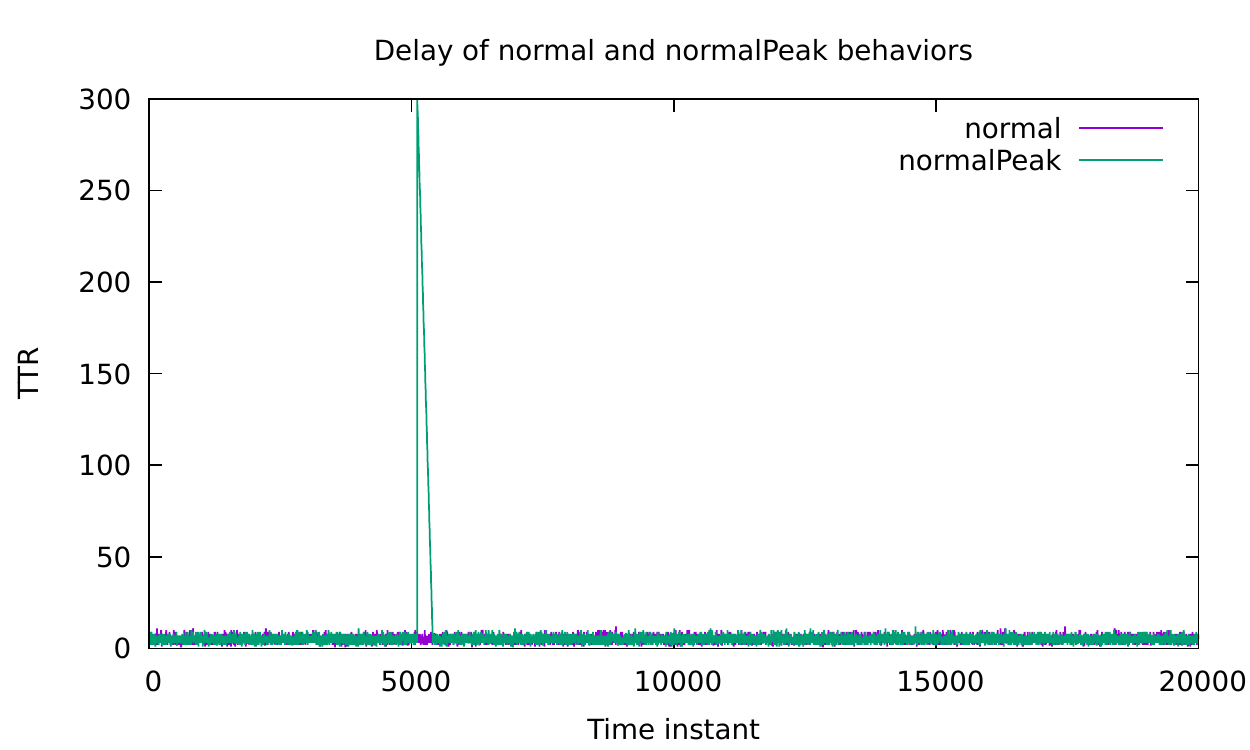}\\
  \multicolumn{1}{l}{\begin{tabular}{l}\\[-0.5em](b) Normal and NormalPeak \end{tabular}}\\
\end{tabular}
  \caption{Examples of network behaviors}
  \label{plot:netbehaviors}
\end{figure}

We can observe an example of the delays observed under these behaviors
in Figure~\ref{plot:netbehaviors}.
%
% In Figure~\ref{plot:netbehaviors} we can also observe the system time
% that gets monitored with a trace of a certain length taking into
% account the sampling period.
%
The system under observation is sampled periodicly, obtaining the
input traces for each of the variables measured.
Thus, having the length of the trace and the sampling period we can
obtain the system time that gets monitored throughout the experiment.
For example, a trace of length $200k$ with a sample period of $30$
seconds, corresponds to monitoring a system during $\approx 2.31$
months.
For some of the experiments the traces of real data available in the
datasets were not sufficiently long, so we extended those traces by
repeating the samples as much as needed to reach the desired trace
length.
Also, some of those traces required interpolation in order to use a
common clock tick for all events, since some of those traces were
based on events instead of sensing periodically a variable.
We did this interpolation whenever needed.

\subsection{Hypothesis}
For the empirical evaluation of this paper we intend to evaluate the
following hypothesis:
\begin{compactitem}
\item (H1) Our time Asynchronous algorithm behaves no worse than the
  synchronous algorithm from~\cite{danielsson19decentralized} when the
  network presents a synchronous behavior.
\item (H2) Synchronous SRV can simulate the monitoring of a time
  asynchronous network with a software layer that provides the
  illusion of synchronicity, but at a very high cost in delays and
  memory usage.
\item (H3) Our theoretical results of
  Section~\ref{sec:resources-eager} hold for the execution of the
  experiments.
\item (H4) Local memory of the root monitor is bounded, resulting in a
  trace length independent monitor our theoretical results predict.
\item (H5) Our algorithm scales in terms of number of monitors-network
  usage. We expect that memory will increase linearly with ``network
  usage'' but will remain constant when increasing the number of local
  monitors. Here we refer with local monitor to a non-empty set of
  streams that are computed at the same network node.
\item (H6) We can benefit from using redundant specifications and
  redundant topologies (exploiting simplifiers) to reduce
  $\TTR$\textit{s} by avoiding delays of slow or faulty links.
\end{compactitem}
\subsection{Empirical Results}
%
% \summary{show tables or plots for each hypothesis that confirm them}
%
In order to validate hypothesis (H1) we built the following
experiments:
\begin{itemize}
\item SmartPolitechDistr: we detect fire hazards by analysing the
  levels of temperature, $\text{CO}_2$ and humidity in the air in
  different rooms in university buildings.
  We use a quantitative robust specification.
\item tomskHeating: we check that the heating system is behaving as
  expected (extracted from the data). Again, this is a quantitative
  robust specification.
\item orange4Home: we detect fire hazards by analysing the activities
  performed by the tenant in the apartment.
\item contextAct: we detect fire hazards by analysing the levels of
  temperature, $\text{CO}_2$ and humidity in the air in different
  rooms in an smart apartment.
  This is also a quantitative specification.
\end{itemize}
Figure~\ref{fig:tadLola-ttr-analysis} shows metrics of the delay of
the root of the specification for the different datasets analyzed with
different network behaviors.
This proves empirically that TADSRV subsumes DSRV with no additional
loss of performance, as expected by our theoretical proofs.
Therefore (H1) holds.
All these different network behaviors show that TADSRV is more general
than DSRV, as we expected.

\begin{figure}[tbh!]
  \centering
  \includegraphics[scale=0.2]{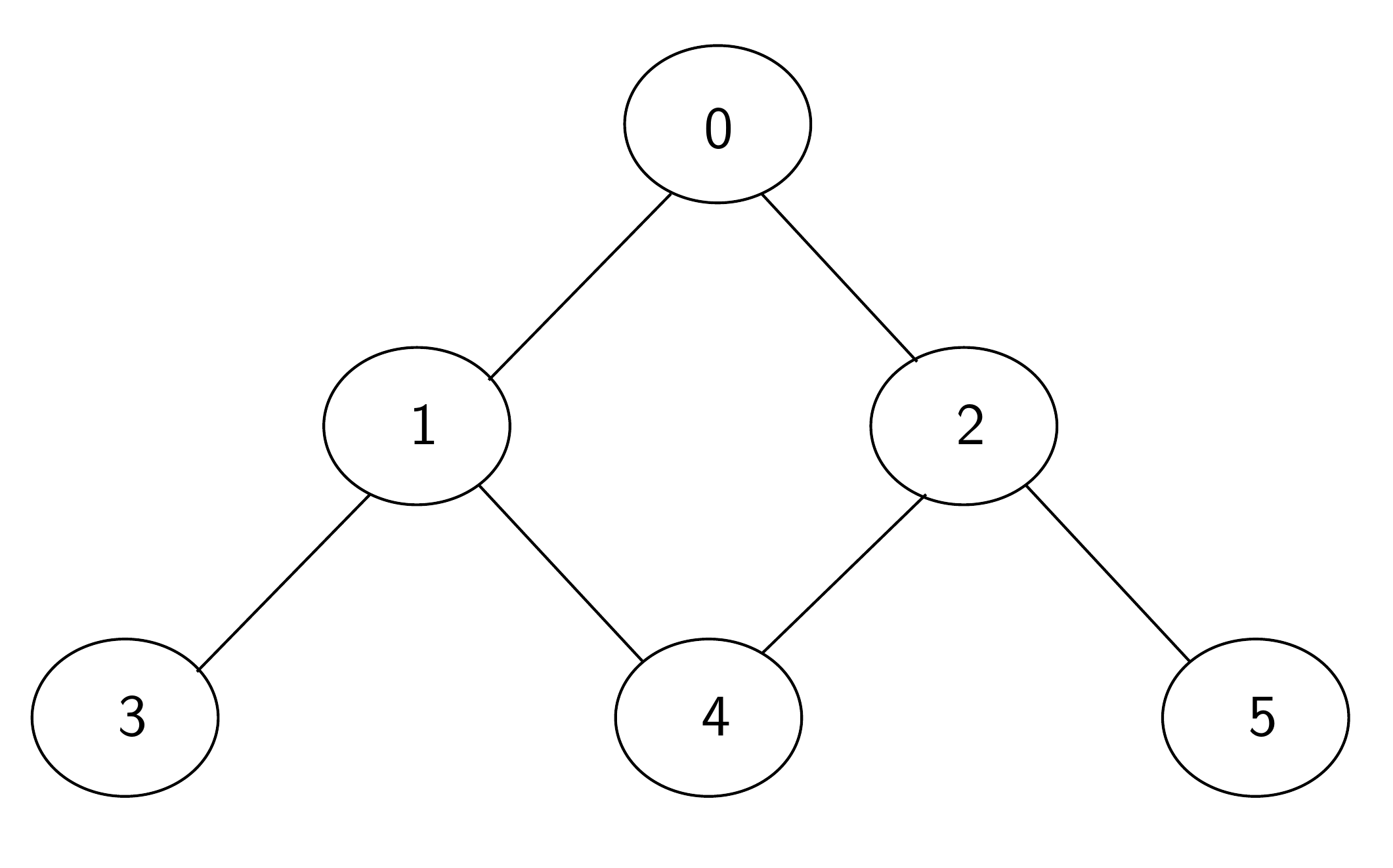} \\
  \caption{Monitor topology of the experiment for (H2)}
  \label{plot:syncSimAsynTopo}
\end{figure}

For the validation of (H2) we built an experiment with the
specification of obtaining both the maximum and sum of the inputs.
We placed this in the topology shown in Figure~\ref{plot:syncSimAsynTopo}.

\begin{figure}[tbh!]
  \begin{tabular}{l}
  \includegraphics[scale=0.6]{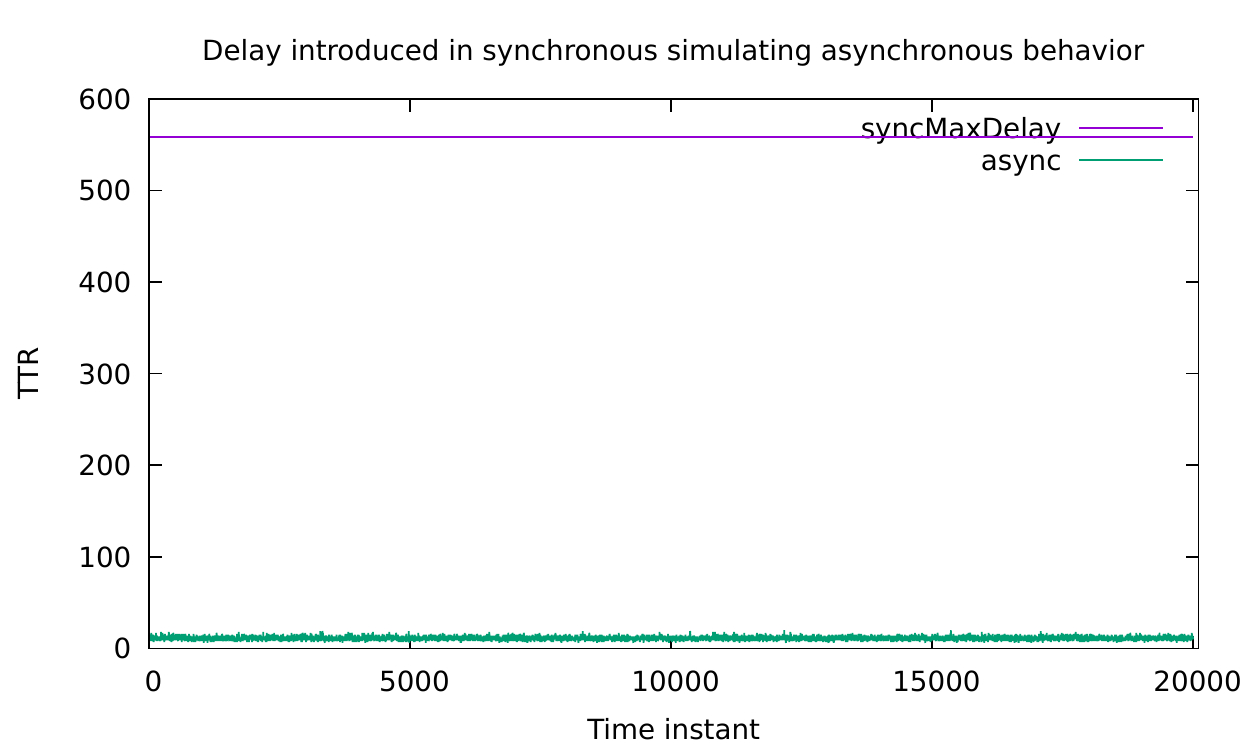} \\
  \includegraphics[scale=0.6]{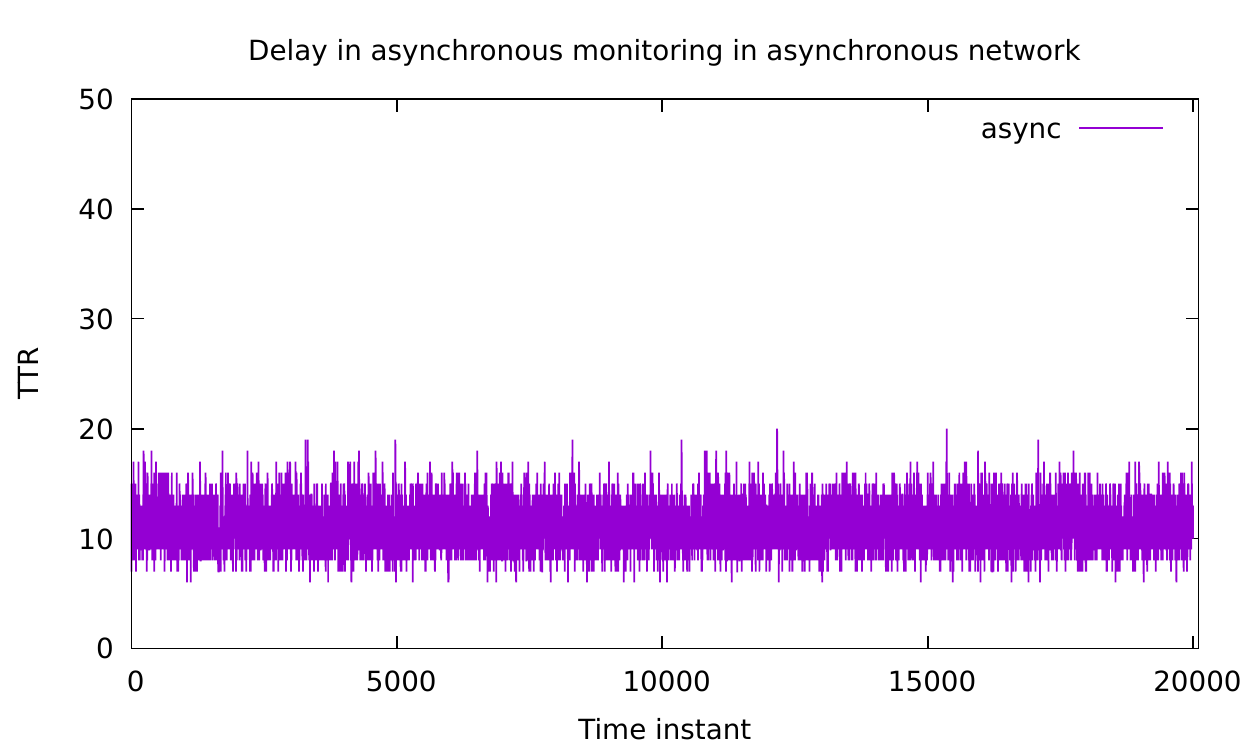} \\
  \multicolumn{1}{l}{\begin{tabular}{l}\\[-0.5em](a) TTR of Synchronous and Asynchronous \end{tabular}} \\
  \includegraphics[scale=0.6]{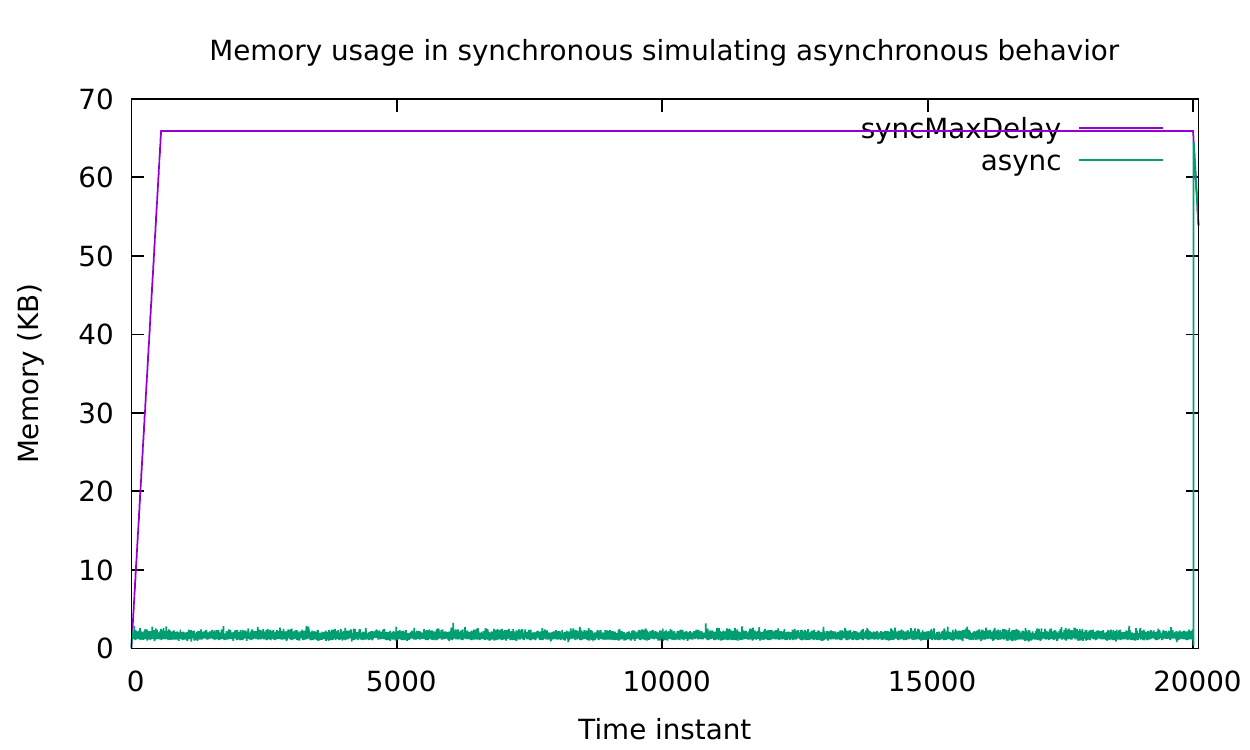} \\
  \includegraphics[scale=0.6]{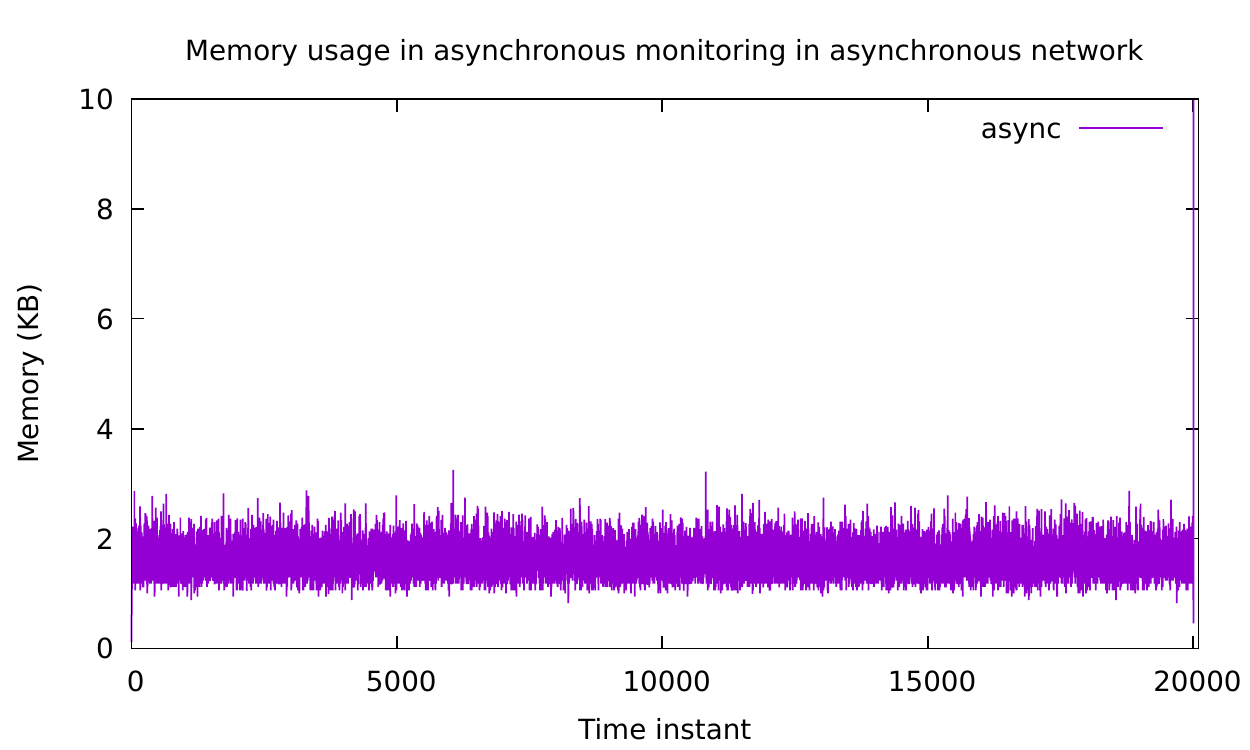} \\
  \multicolumn{1}{l}{\begin{tabular}{l}\\[-0.5em](b) Memory of root monitor of synchronous and asynchronous \end{tabular}}\\
\end{tabular}
  \caption{Synchronous and asynchronous in an asynchronous network
    with details of asynchronous}
  \label{plot:synSimAsynMTR}
\end{figure}

%
%NOT NOWBut in order to simulate the synchronous monitoring we forced the monitors to wait until they have all the inputs for a certain tick before computing any instant stream in $\U$.
We looked for the maximum delay present in the normalPeak traces that
we have and used that duration as the global delay between each pair
of monitors in the synchronous scenario.
We measured both settings: simulating synchronicity and the execution
of the timed asynchronous algorithm.
The results are shown in Figure~\ref{plot:synSimAsynMTR}.
The figure shows that we can emulate TADSRV with DSRV but with a high
cost in memory usage ($+200\%$ than the worst instant) and incurring
in delays of
$\text{\textit{worst delay}} * \text{\textit{depth of topology}}$,
which in this case is $558$ instants.
This corresponds to an increase of around $30$ times the delay of the
timed asynchronous.
Therefore, (H2) holds as well.
This results makes it clear that it is not feasible in practice to use
DSRV in a time asynchronous scenario (even with the layer that
simulates synchrony), where the contribution of this work applies
naturally with much better performance.
%
%\begin{figure}[tbh!]
%  \begin{tabular}{cc}
%  \label{plot:synSimAsynMTR}
%  %\includegraphics[scale=0.7]{plots/smartPolitechDistr_const_tl_2000_memoryAvg-crop}
%  %\includegraphics[scale=0.7]{plots/syncSimulatingAsyn_normalPeak_tl_20000_memoryAnalysis-crop}
%  %\includegraphics[scale=0.37]{plots/syncSimulatingAsyn_synMaxDelayMemGP}
%
%  \end{tabular}
%  \caption{Maximum memory of root monitor of synchronous and asynchronous in an asynchronous network}
%  \label{plot:synSimAsynMemory}
%\end{figure}

Also, we can see that the TTRs obtained empirically are below or equal
to our estimated bounds calculated a-priori with the equations
described in Section~\ref{sec:resources-eager}.
Hence, (H3) holds.

For the validation of (H4)---studying the scalability in terms of
trace length---we used the smartPolitechDistr dataset and run it with
a trace of $200k$ instants with the normalPeak behavior.
%
% Here we used the specification of the SmartPolitech experiment.
%
In the extract shown below we compute both a Boolean and a
quantitative stream to look for temperature uprisings.
%(Actually it is parameterized by the monitor in which it is computed, since we need to compute this with the data of each room and we model a monitor per room; factors are extracted from data analysis)

\begin{figure}[tbh!]
  \centering
  \includegraphics[scale=0.6]{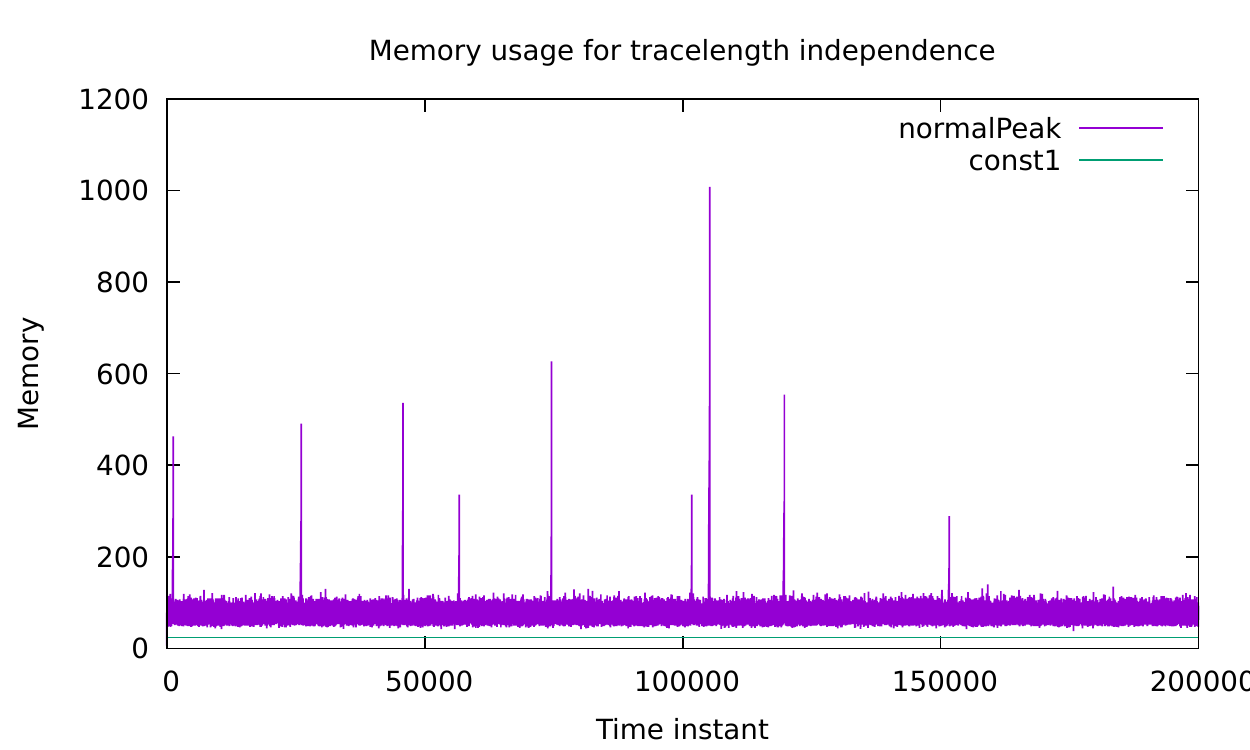}
  \caption{Maximum memory of root monitor of the last 10000 instants}
  \label{plot:tracelen-indep}
\end{figure}

{\footnotesize\begin{lstlisting}[language=LOLA]
  define bool temp_up eval =
    temp > 1.1 * tempini and temp <= 1.6 * tempini
  define num temp_up_q eval =
    if temp <= 1.1*tempini then 0 else
    if temp > 1.6*tempini then 1 else
    (temp - 1.1*tempini)/(1.6*tempini-1.1*tempini)
  define bool temp_spike eval =
    temp > 1.6 *tempini
  define num temp_spike_q eval =
    if temp <= 1.6*tempini then 0 else
    if temp > 2 * tempini then 1 else
    (temp - 1.6*tempini)/(2*tempini-1.6*tempini)
\end{lstlisting}}
Figure~\ref{plot:tracelen-indep} shows that the memory used in the
root monitor of this experiment remains bounded.
The pikes in memory correspond to higher delays in the network links
among nodes.
This forces monitors to keep records in their memory until the
messages that they need arrive, allowing the monitor to resolve
streams and prune their memories.
This result suggests that the algorithm with a decentralized
efficiently monitorable specification can behave in a trace-length
independent fashion, validating hypothesis (H4).

\begin{figure}[tbh!]
  \centering
  \includegraphics[scale=0.6]{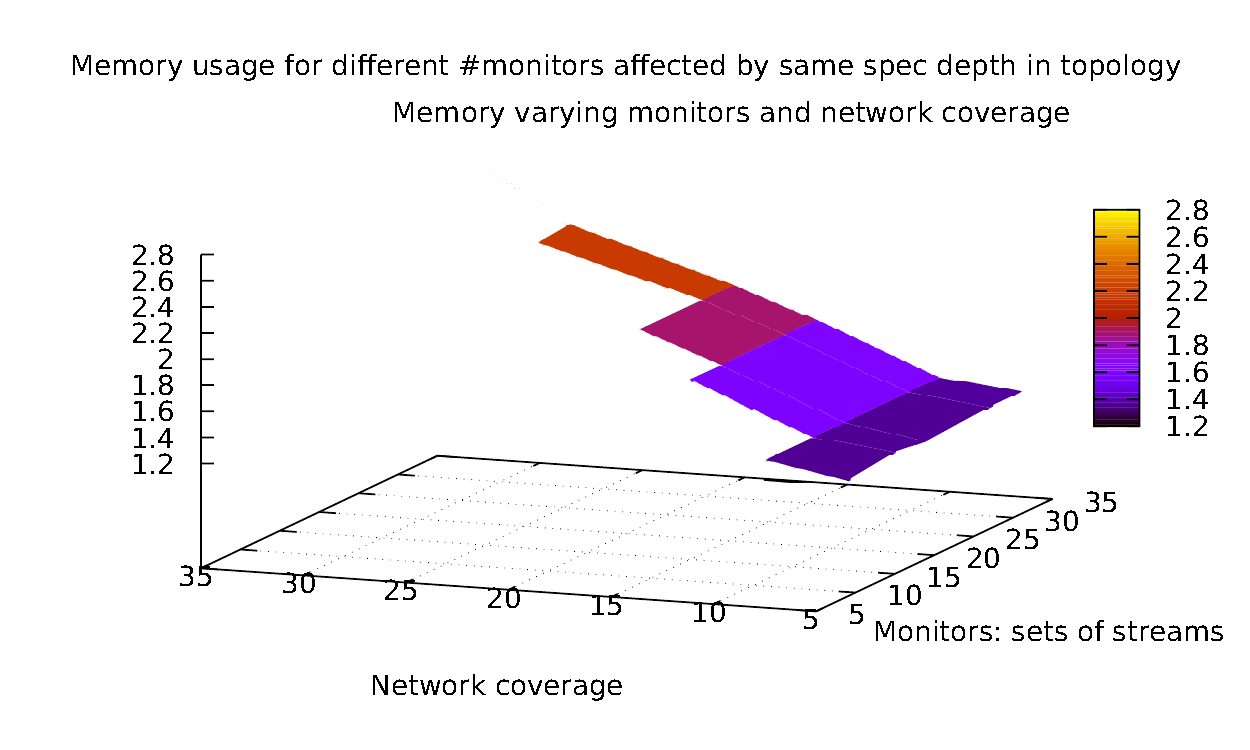}
  \caption{Average memory usage of the monitor that uses most memory
    of the last 1000 instants, tlen 20k, with different network
    coverages}
  \label{plot:monitor-scalability}
\end{figure}

Figure~\ref{plot:monitor-scalability} shows that the memory usage of a
single monitor does not depend on the number of other monitors in the
network but it depends on the maximum depth of its specification that
travels the network.
%
% For this experiment we consider that a ``monitor'' is a non-empty set
% of streams that are computed at the same network node.
%
In this experiment the depth of the specification deployed in the
network was kept constant ($5$) while we changed the number of
monitors in a binary tree topology (preserving the depth in one
branch).
%
%We obtained the expected results of no memory variation as
%shown in Figure~\ref{plot:scalability-monitors-nodes}.
%
The intuition is that the variable that affects memory usage is not
how many monitors we have but the number of network nodes and links
among them that affect the monitoring performance.
This is because the more links, the higher the probability that a
failure in the network (modelled as a delay) affects the run.
These results prove that hypothesis (H5) holds.
\subsubsection{Redundancy and Delays.}
\begin{figure}[tbh!]
  \centering
  \includegraphics[scale=0.6]{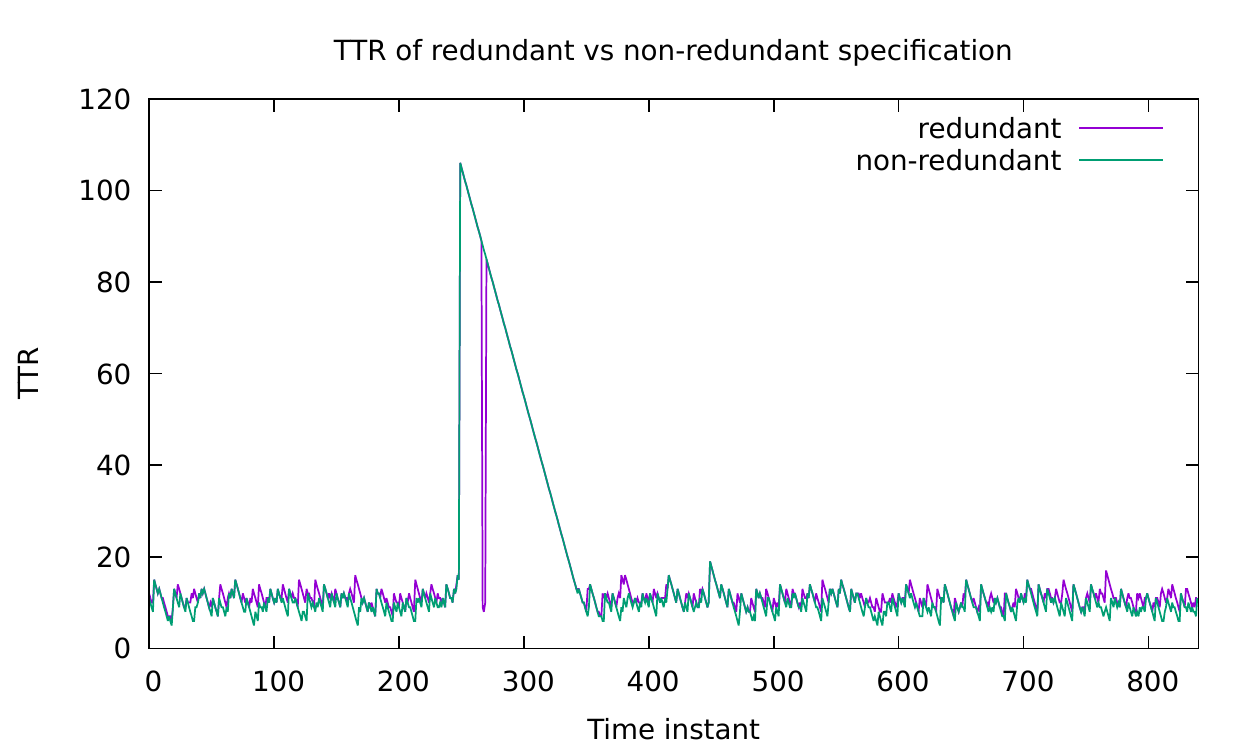}
  \caption{Benefits of using Redundancy in terms of accumulated
    delays}
  \label{plot:redundancy-delays}
\end{figure}

% \summary{define what we call redundancy in spec and in topo, the
%  objective that we have in terms of performance, spec and show plot}
%\TableTopology
%
In this subsection we take a closer look at hypothesis (H6), so we
build the topology and the specification to minimize the
$\TTR$\textit{s} of the instant variables.
We seek to benefit from using simplifiers to minimize the effect
network delays of messages required to compute the instant variables.
Thus, we intend to exploit the messages that go through the fastest
path in the network from the nodes that read the inputs to the nodes
that compute the root of the specification.
Intermediate results are generated faster in the least congested
deployment and messages will travel through the least weight path (in
terms of accumulated delays) between the inputs and the root of the
specification yielding a minimum TTR for the instant variables.
This improvement can be achieved because intermediate results from
slower monitors will not be needed due to the use of simplifiers, and
therefore the engine will not wait to achieved a final result of the
root monitor.
We build the following fragment of the specification for the data in
smartPolitech, where we make the streams \verb+C3_fire_risk_q+ and
\verb+C3_fire_risk_q_red+ redundant of each other and we deploy them
in different monitors so that they are affected by different
delays.
We use a normal delay for the whole network but introduce a failure in
the form of a peak in the delays between the monitors connected to
monitor 3.
This will make the path through monitor 2 faster.
We can observe in Figure~\ref{plot:redundancy-delays} how the delay of
obtaining the value for the root of the specification takes the best
delay possible.
Since we use an OR to take advantage of the symplifiers, in the best
case verdict (outcome true) there is a gain, but in the worst case
verdict (false) the redundant solution gains no speed as the
engine needs to wait for all the values to calculate the OR.

{\footnotesize\begin{lstlisting}[language=LOLA]
@0{
define bool C3_alarm eval =
  (C3_fire_risk or C3_fire_risk_red) and
  (C3_fire_risk_q > 0.5 or C3_fire_risk_q_red > 0.5)
}
@2{
define bool C3_fire_risk_red eval =
  AND(C3_temp_spike,C3_co2_spike,C3_humid_down)
define num C3_fire_risk_q_red eval =
  AVG(C3_temp_spike_q,C3_co2_spike_q,C3_humid_down_q)
}
@3{
define bool C3_fire_risk eval =
  AND(C3_temp_spike,C3_co2_spike,C3_humid_down)
define num C3_fire_risk_q eval =
  AVG(C3_temp_spike_q,C3_co2_spike_q,C3_humid_down_q)
}
\end{lstlisting}}
Figure~\ref{plot:redundancy-delays} shows the difference between using
the redundant specification with redundant topology and not using any
redundancy.
Even though a general study of exploiting redundant paths in the
network is out of the scope of this paper, this case study illustrates
how redundant deployments can improve decentralized monitoring.

%%% Local Variables:
%%% TeX-master: "main.tex"
%%% TeX-PDF-mode: t
%%% End:

\section{Lazy Algorithm}
\label{sec:lazy}
%I use the suffix -complete for any label that refers to eager and lazy

We introduce now a variant of Algorithm~\ref{alg:local-algo-tadLola}
where some of the streams are not sent unless their values are requested.
This is beneficial in cases where their value is rarely needed.
We call these \emph{lazy streams}.

To introduce the modified algorithm we need to introduce a new type of
message: the request message.
We also call a response message to the messages containing the value
of an instant variable.

\begin{itemize}
%REQUESTS ARE USED FOR LAZY ONLY, see LAZY section in efficient.tex
\item \textbf{Response} messages: $(\RESP,\InstVar{s}{k},c, n_s,n_d)$
  where $\InstVar{s}{k}$ is an instant variable, $c$ is a constant of
  the same datatype as $\InstVar{s}{k}$, $n_s$ is the source node and
  $n_d$ is the destination node of the message.
\item \textbf{Requests} messages: $(\REQ,\InstVar{s}{k},n_s,n_d)$
  where $\InstVar{s}{k}$ is an instant variable, $n_s$ is the source
  node and $n_d$ is the destination node of the message.
\end{itemize}
Again, if $\MSG=(\REQ,\InstVar{s}{k},n_s,n_d)$, then $\MSG.\SRC=n_s$,
$\MSG.\DST=n_d$, $\MSG.\TYPE=\REQ$, $\MSG.\STREAM=\InstVar{s}{k}$.
Similarly, for a response message we have the same, the only
difference is that we add $\MSG.\VAL=c$.
%
%\subsubsection{Stream Assignment and Communication Strategy}

Each stream variable $v$ can be assigned one of the following two
\emph{communication strategies} to denote whether an instant value
$\InstVar{v}{k}$ is automatically communicated to all potentially
interested nodes, or whether its value is provided upon request only.
Let $v$ and $u$ be two stream variables such that $v$ appears in the
equation of $u$ and let $n_v=\mu(v)$ and $n_u=\mu(u)$.
\begin{itemize}
\item \textbf{Eager communication}: the node $n_v$ informs $n_u$ of
  every value $\InstVar{v}{k}=c$ that it resolves by sending a message
  $(\RESP,\InstVar{v}{k},c,n_v,n_u)$.
  This is what we have used previously in the paper.
\item \textbf{Lazy communication}: node $n_u$ requests $n_v$ the value
  of $\InstVar{v}{k}$ (in case $n_u$ needs it to resolve $\InstVar{u}{k'}$ for some
  $k'$) by sending a message $(\REQ,\InstVar{v}{k},n_u,n_v)$. When $n_u$ receives
  this message and resolves $\InstVar{v}{k}$ to a value $c$, $n_u$ will respond
  with $(\RESP,\InstVar{v}{k},c,n_v,n_u)$.
\end{itemize}
Each stream variable can be independently declared as eager or lazy.
We use two predicates $\ISEAGER(u)$ and $\ISLAZY(u)$ (which is defined
as $\neg \ISEAGER(u)$) to indicate the communication strategy of
stream variable $u$.
Note that the lazy strategy involves two messages and the eager
strategy only one, but eager sends every instant variable resolved,
while lazy will only sends those that are requested.
In case the values are almost always needed, eager is preferable while
if values are less frequently required lazy is preferred.
We now need to add the communication strategy to the definition of the
decentralized SRV problem.
%
%\begin{definition}
A decentralized SRV problem $\tupleof{\varphi,\topo,\mu, \ISEAGER}$ is
now characterized by a specification $\varphi$, a topology $\topo$, a
stream assignment $\mu$ and a communication strategy for every stream
variable.
%\end{definition}

\subsection{Lazy DSRV Algorithm for Timed Asynchronous Networks}
We extend our local monitor to $\tupleof{\QUEUE_n, U_n, R_n, \PEN_n,\WAIT_n}$
adding the following two storages:
\begin{itemize}
\item \textbf{Pending} requests $\PEN_n$, where $n$ records instant
  variables that have been requested from $n$ by other monitors but that
  $n$ has not resolved yet.
\item \textbf{Waiting} for responses $\WAIT_n$, where $n$ records
  instant variables that $n$ has requested from other nodes but has
  received no response yet.
\end{itemize}
The storage $\WAIT_n$ is used to prevent $n$ from requesting the same
value twice while waiting for the first request to be responded.
An entry in $\WAIT_n$ is removed when the value is received, since the
value will be subsequently fetched directly from $R_n$ and not
requested through the network.
The storage $\PEN_n$ is used to record that a value that $n$ is
responsible for has been requested, but $n$ does not know the answer
yet.
When $n$ computes the answer, then $n$ will send the corresponding response
message and remove the entry from $\PEN_n$.
Finally, request messages are generated for unresolved lazy instant variables
and inserted in the queues of the corresponding neighbors.

\AlgorithmTADLola

More concretely, every node $n$ will execute the procedure \Monitor\
shown in Algorithm~\ref{alg:local-algo-tadLola}, which invokes \Step\ in every
clock tick until the input terminates or ad infinitum.
Procedure \Finalize\ is used to resolve the pending values at the end
of the trace to their default if the trace ends.
Procedure \Step\ now executes some modified procedures and additional
steps:
\begin{enumerate}
\item \textbf{Process Messages}:
  %Lines $7$ invokes \ProcessMessages procedure
  %in lines $23$-$28$ that deal with the processing of incoming messages.
%
%  First, Lines $13$-$14$ route messages with a different destination.
%
  Line $26$ annotates requests in $\PEN_n$, which will be later
  resolved and responded.
  Lines $27$-$28$ handle response arrivals, adding them to $R_n$ and
  removing them from $\WAIT_n$.
\item \textbf{Send Responses:} %Line $11$ invokes \SendResponses, in lines $29$-$36$.
  %First lines $30$-$32$ send messages for all newly resolved eager variables.
  %
  Lines $33$-$36$ deal with pending lazy variables.
  If a pending instant variable is now resolved, the response message
  is sent and the entry is removed from $\PEN_n$.
\item \textbf{Send new Requests:} Lines $37$-$41$ send new request
  messages for all lazy instant streams that are now needed.
\item \textbf{Prune:} Line $42$-$44$ prunes the set $R$ from information
  that is no longer needed. See section~\ref{sec:lazy-prune}.
\end{enumerate}

\subsection{Formal Correctness}
\label{subsec:correctness-complete}
% \summary{formal proof of convergence and correctness}

We now show that our solution is correct again by proving that the
output computed is the same as in the denotational semantics, and that
every output is eventually computed.

\newcounter{thm-convergence-complete}
\setcounter{thm-convergence-complete}{\value{theorem}}
\begin{theorem}
  \label{thm:convergence-complete}
  All of the following hold for every instant variable $\InstVar{u}{k}$:
  \begin{compactenum}
    \item[\textup{(1)}] The value of $\InstVar{u}{k}$ is eventually resolved.
    \item[\textup{(2)}] The value of $\InstVar{u}{k}$ is $c$ if and only if $(\InstVar{u}{k},c)\in R$ at
      some instant.
    \item[\textup{(3)}] If $\ISEAGER(u)$ then a response message for
      $\InstVar{u}{k}$ is eventually sent.
    \item[\textup{(4)}] If $\ISLAZY(u)$ then all request messages for $\InstVar{u}{k}$
      are eventually responded.

  \end{compactenum}
\end{theorem}
%
%%%% PROOF IN APPENDIX
\begin{proof}
  The proof proceeds by induction in the evaluation graph, showing
simultaneously in the induction step $(1)$-$(4)$ as these depend on
each other (in the previous inductive steps).
  Let $M$ be a length of a computation and $\sigma_I$ be an input of
  length $M$.
  Note that $(1)$ to $(4)$ above are all statements about instant
  variables $\InstVar{u}{k}$, which are the nodes of the evaluation graph
  $G_{\varphi,M}$.
  We proceed by induction on $G_{\varphi,M}$ (which is acyclic because
  $D_\varphi$ is well-formed).
  \begin{itemize}
  \item \textbf{Base case}: The base case are vertices of the
    evaluation graph that have no outgoing edges, which are either
    instant variables that correspond to inputs or to defined
    variables whose instant equation does not contain other instant
    variables.
    Statement $(1)$ follows immediately for inputs because at instant
    $k$, $\InstVar{s}{k}$ is read at node $\mu(k)$.
    For output equations that do not have variables, or whose
    variables have offsets that once instantiated become negative or
    greater than $M$, the value of its leafs is determined either
    immediately or at $M$ when the offset if calculated.
    At this point, the value computed is inserted in $R$, so $(2)$
    also holds at $\mu(u)$.
    Note that $(2)$ also holds for other nodes because the response
    message contains $\InstVar{u}{k}=c$ if and only if $(\InstVar{u}{k},c)\in R_n$, where
    $\mu(u)=n$.
    %
    %If $s$ is eager,
    Then the response message is inserted exactly at
    the point it is resolved, so $(1)$ implies $(3)$.
    Finally, $(4)$ also holds at the time of receiving the request
    message or resolving $\InstVar{u}{k}$ (whatever happens later).
  \item \textbf{Inductive case}: Consider an arbitrary $\InstVar{u}{k}$ in the
    evaluation graph $G_{\varphi,M}$ and let
    $\InstVar{u_1}{k_1}\ldots \InstVar{u_l}{k_l}$ the instant variables that $\InstVar{u}{k}$
    depends on.
    These are nodes in $G_{\varphi,M}$ that are lower than $\InstVar{u}{k}$ so
    the inductive hypothesis applies, and $(1)$-$(4)$ hold for these instant variables.
    Let $n=\mu(u)$.
    At instant $k$, $\InstVar{u}{k}$ is instantiated and inserted in $U_n$. 
    At the end of cycle $k$, lazy variables among $\InstVar{u_1}{k_1}\ldots
    \InstVar{u_l}{k_l}$ are requested.
    By induction hypothesis, at some instant all these requests are
    responded by $(1)$ and $(4)$.
    Similarly, the values of all eager variables are calculated and
    sent as well (by $(1)$ and $(3)$ which hold by IH).
    At the latest time of arrival, the equation for $\InstVar{u}{k}$ has no more
    variables and it is evaluated to a value, so $(1)$ holds and $(2)$
    holds for $\InstVar{u}{k}$ at $n$.
    At this point, if $\ISEAGER(u)$ then the response message is sent
    (so $(1)$ holds for $\InstVar{u}{k}$) and if $\ISLAZY(u)$ then all requests
    (previously received in $\PEN_n$ or future requests) are answered,
    so $(1)$ also holds.
  \end{itemize}
  This finishes the proof.
\end{proof}

\subsection{Resources for Lazy}

Analyzing the lazy case requires modifications.
%
% First of all, we note that we need to replace this notion that we had
% in synchronous DSRV~\cite{danielsson19decentralized}, that we could
% count time and infer that after a certain instant no request will ever
% arrive for an instant variable because it would already had arrived.
%
In timed asynchronous networks we need to introduce a new kind of
message to provide \emph{confirmations} that are only used to inform
the receiving node that some instant variables are not needed so they
can be pruned.
This new message have the following form:
\begin{itemize}
\item \textbf{Confirmation} messages:
  $(\CONFIRM,\InstVar{s}{k},n_s,n_d)$ where $\InstVar{s}{k}$ is an
  instant variable, $n_s$ is the source node and $n_d$ is the
  destination node of the message.
\end{itemize}
This message will be interpreted as the source node $n_s$ has resolved
instant variables $s$ up to $k$.
This information allows the destination node to conclude that instant
variables required at the remote node for nodes that have been
resolved are no longer necessary.
We change $\MTRToRem{\MTR}$ to include that the response gets emitted
when the request arrives or when the remote instant variable gets
resolved, whichever happens later.
\begin{align*}
  & \MTRToRem{\MTRlazy}(\InstVar{s}{t}) \DefinedAs \max(t, M(\InstVar{s}{t})\\
  \intertext{where}
  & M(\InstVar{s}{t})=\{\ArrivalTime{r}{s}(t') ~\text{s.t.} 
   \depWeightSub{s}{w}{rem}{r} \text{ and } t+w \geq 0 \} ) \\
  \intertext{and}
& t' = \max(\ArrivalTime{s}{r}(t),\MTRlazy(\InstVar{r}{t+w})) 
\end{align*}
Here $\ArrivalTime{s}{r}(t)$ is the time when the request is sent,
that is, when the instant variable $s$ gets instantiated and stored in
$U$.
$\MTRToRem{\MTRlazy}(\InstVar{r}{t+w})$ is when the remote instant
stream gets resolved.
Finally
$\ArrivalTime{r}{s}(t')$
is the moment at which the response of the lazy instant stream
variable arrives at the requesting node.

\subsubsection{Instantiation to Synchronous}
Again, we first consider the case where the delay of any link to be a
constant throughout the execution.
This constant is useful to simplify the equations but we need to
consider now that for each instant variable we need a request and
afterwards a response, in order to get the remote value.
%
% The synchronous model of computation is a particular case of the
% timed-asynchronous model where all message delays between two monitors
% take exactly the same amount of time, a constant that is known a
% priori.
%
Again, $\DIST{r}{s}$ is used to represent the delay that every message
will take from $\mu(r)$ to $\mu(s)$, independently of the time instant
at which the message is sent.
We use this knowledge to simplify $\MTRToRem{\MTRlazy}$ for
synchronous networks as follows
%
%\begin{align*}
%  &\MTRsynrem(\InstVar{s}{t}) \DefinedAs \max(t, \\
%  &\{ \MTRsyn(\InstVar{r}{t+w}) + \DIST{r}{s} \;|\;\depWeightSub{s}{w}{rem}{r}, t+w \geq 0 \})
%\end{align*}
%
\begin{align*}
  & \MTRToRem{\MTRsynlazy}(\InstVar{s}{t}) \DefinedAs \max(t, t') ~\text{s.t.} \depWeightSub{s}{w}{rem}{r}; t+w \geq 0 \}) \\
  \intertext{where}
& t' = \{ \DIST{r}{s} + \max(t + \DIST{s}{r},\MTRsynlazy(\InstVar{r}{t+w})) 
\end{align*}

Where the value of the remote instant variable arrives when the
response message arrives $\DIST{r}{s}$, which is emitted either when
the request arrived $t + \DIST{s}{r}$ or when the remote value is
resolved $\MTRsyn(\InstVar{r}{t+w})$, whichever ocurrs later.

\subsubsection{ \GlobalTimeAdv Bounded delays}

Now we consider that case where we know a maximum delay in the network
that upper bounds all the other delays in the network behavior.
%
% We can consider this as the delay that is affecting each link in every
% time instant and substitute in the equations, obtaining a upper bound
% and thus, the moment at which every instant variable is guaranteed to
% be resolved.
% %
% An execution of timed-asynchronous monitors has \globalTimeAdv bounded
% delays if there is an upper bound on the arrival time of every message ever sent.
%
%\cesartext{Show an example of network behavior that is not globally
%  bounded}
%
%Formally, if there is a $d$ such that for every streams $r,s$ and for
%every time $t$, $\ArrivalTime{r}{s}(t)\leq t+d$.
%
Substituting the upper-bound value $d$ in the equations for $\MTR$, we
obtain an constant upper-bound on the $\MTR$ (although this value can
be a gross over-approximation):
%
%\begin{align*}
%  &\MTRgrem (\InstVar{s}{t}) \leq \max(t,\{\\
%  &\hspace{4em} \MTRg(\InstVar{r}{t+w}) + d \;|\;\depWeightSub{s}{w}{rem}{r}, t+w \geq 0 \})
%\end{align*}
%
\begin{align*}
  & \MTRToRem{\MTRglazy}(\InstVar{s}{t}) \DefinedAs \max(t, M(\InstVar{s}{t}))\\
  \intertext{where}
  & M(\InstVar{s}{t}=\{ d + \max(t + d,t') ~\text{s.t.}~\depWeightSub{s}{w}{rem}{r}; t+w \geq 0 \}) \\
  \intertext{and}
& t' = \MTRglazy(\InstVar{r}{t+w})
\end{align*}

\subsubsection{ \LocalTimeAdv Bounded delays}

Finally, we do not assume an \emph{\globalTime} bound on the delays of
the network.
Instead, we can just look at what affects the computation of the
instant variables, that is, other instant variables that it depends on
and the network delays that affect the messages to compute those
instant variables.
We take into account again the window $\Win(\InstVar{s}{t})$, which
contains the interval that includes all the instants at which values
that influence $\InstVar{s}{t}$ are resolved and sent.
This window always ends at most at $\MTR(\InstVar{s}{t})$.
Inside this window we can find the worst delay of a message sent for
the computing of the instant variable: $\dworst(\InstVar{s}{t})$.
Then, we can bound $\MTR(\InstVar{s}{t})$ as follows for the lazy
case:

%\begin{align*}
%  &\MTRToRem{\MTRlocallazy}(\InstVar{s}{t}) \leq \max(t,\\
%  &\{\MTRlocallazy(\InstVar{r}{t+w})+\dworst(\InstVar{s}{t})\;|\;\wdirrem{s}{w}{r} \}).
%\end{align*}
%
\begin{align*}
& \MTRToRem{\MTRlocallazy}(\InstVar{s}{t}) \DefinedAs \max(t, \{t'~\text{s.t.} \depWeightSub{s}{w}{rem}{r} ; t+w \geq 0 \})\\
& \text{where} \\
& t' = \dworst(\InstVar{s}{t}) + \max\binom{\dworst(\InstVar{s}{t})}{\MTRlocallazy(\InstVar{r}{t+w})} 
\end{align*}

Here, $\dworst(\InstVar{s}{t})$ is the time for worst message
affecting the computation of $\InstVar{s}{t}$, so the window for
obtaining this value considers both request and response messages.
We use this value to bound both the request and the response.
First, we obtain the latter instant at which either the request arrives or
the remote dependency is resolved in
$\max(\dworst(\InstVar{s}{t}),\MTRlocallazy(\InstVar{r}{t+w}))$ and
then we add the time for the response message to arrive with the value
in $\dworst(\InstVar{s}{t})$.
Obtaining the moment at which we know that the remote dependency is guaranteed
to be resolved and its value arrived at the requesting network node.

\subsection{Pruning the Resolved Storage.}
\label{sec:lazy-prune}

We are finally ready to prune $R_n$ for the lazy algorithm case
because we know now when every instant variable will be resolved.
\begin{theorem}
  \label{thm:mtr-lazy}
  Every unresolved instant variable $\InstVar{s}{t}$ that is lazy in $U_n$ is
  resolved at most at $\MTRlazy(\InstVar{s}{t})$.
\end{theorem}
As soon as this moment is reached, considering that the network delays
are bounded, a confirmation message will be sent to those monitors
where lazy instant variables that are dependencies to the resolved
instant variable are computed and this message will arrive in bounded
time.
Then the receiving node can prune the corresponding instant variables
from its memory.
Now we need to add $\tconf$ in this theorem which is the time for the
confirmation message to arrive:

Every unresolved $\InstVar{s}{k}=e$ in $U_n$ is pruned at most at
$max(\{\MTRlazy(\InstVar{u}{k-w})+\tconf_u\})$.  Where
$\InstVar{u}{k-w}$ is a remote instant variable that contains
$\InstVar{s}{k}$ in its equation and $\tconf_u$ is the time for the
confirmation message to travel from $\mu(u)$ to $\mu(s)$ sent at time
$\MTRlazy(\InstVar{u}{k-w})$.  % Then at time the confirmation
This message arrives at destination in bounded time and the instant
variable gets pruned.
Because at that point the receiving node knows that the instant
variable is no longer needed and can prune it even if it is not
resolved yet.
With this mechanism, we can assure that every instant variable will be
in memory ($U_n, R_n$) for a bounded amount of time.
This implies that decentralized efficiently monitorable specifications
in timed asynchronous networks can be monitored with bounded
resources.
The bound depends only linearly on the size of the specification, the
diameter of the network and the delays among the nodes of the network.

%%% Local Variables:
%%% TeX-master: "main.tex"
%%% TeX-PDF-mode: t
%%% End:

\section{Conclusions and Future Work}
\label{sec:conclusions}
% \summary{contribs: tadlola algo, formal proofs, conditions for
% bounded memory, implementation, empirical eval, redundancy
% technique}

We have studied the problem of decentralized stream runtime
verification for timed asynchronous networks where messages can take
an arbitrary ammount of time to arrive.
This problems starts from a specification and a network.
Our solution consists of a placement of output streams and an online
local monitoring algorithm that runs on every node.
We prove the termination and correctness of the proposed algorithm.
We have captured specifications and network assumptions (synchronous, \globalTime and \localTime bounds) that guarantee
that the monitoring can be performed with constant memory
independently of the length of the trace showing that our solution subsumes
the previous synchronous algorithm.
We report on an empirical evaluation of our prototype tool tadLola.
Our empirical evaluation shows that placement is crucial for
performance and suggest that in most cases careful placement can lead
to bounded costs and delays.
As future work we plan to extend our solution to disaster scenarios
where some links may present a delay ad infinitum, so no message can
traverse that link.
Our intuition is that we could use redundancy in the specifications
and the network topology to provide resilience against faulty network
links while also providing better performance than just by replicating
the time asynchronous algorithm and running them in parallel isolated
from each other.

%%% Local Variables:
%%% TeX-master: "main.tex"
%%% TeX-PDF-mode: t
%%% End:

%\vfill
%\pagebreak
%\appendix
%\input{appendix}

%%% Local Variables:
%%% TeX-PDF-mode: t
%%% End:

%\vfill
%\pagebreak

%\addcontentsline{toc}{chapter}{Bibliography}
\bibliographystyle{apalike}
\bibliography{papers}

\begin{thebibliography}{}

\bibitem[Asarin et~al., 2002]{TRE}
Asarin, E., Caspi, P., and Maler, O. (2002).
\newblock Timed regular expressions.
\newblock {\em J. {ACM}}, 49(2):172--206.

\bibitem[Barringer et~al., 2004]{barringer04rule}
Barringer, H., Goldberg, A., Havelund, K., and Sen, K. (2004).
\newblock Rule-based runtime verification.
\newblock In {\em Proc. of the 5th Int'l Conf. on Verification, Model Checking
  and Abstract Interpretation (VMCAI'04)}, volume 2937 of {\em LNCS}, pages
  44--57. Springer.

\bibitem[Basin et~al., 2015]{basin15failure}
Basin, D., Klaedtke, F., and Zalinescu, E. (2015).
\newblock Failure-aware runtime verification of distributed systems.
\newblock In {\em Proc. of the 35th IARCS Annual Conf on Foundations of
  Software Technology and Theoretical Computer Science (FSTTCS'15)}, volume~45
  of {\em LIPIcs}, pages 590--603. Schloss Dagstuhl--Leibniz-Zentrum fuer
  Informatik.

\bibitem[Bauer et~al., 2013]{bauer13propositional}
Bauer, A., K{\"{u}}ster, J., and Vegliach, G. (2013).
\newblock From propositional to first-order monitoring.
\newblock In Legay, A. and Bensalem, S., editors, {\em Runtime Verification -
  4th International Conference, {RV} 2013, Rennes, France, September 24-27,
  2013. Proceedings}, volume 8174 of {\em Lecture Notes in Computer Science},
  pages 59--75. Springer.

\bibitem[Bauer et~al., 2011]{bauer11runtime}
Bauer, A., Leucker, M., and Schallhart, C. (2011).
\newblock Runtime verification for {LTL} and {TLTL}.
\newblock {\em ACM Transactions on Software Engineering and Methodology},
  20(4):14.

\bibitem[Bauer and Falcone, 2012]{bauer12decentralised}
Bauer, A.~K. and Falcone, Y. (2012).
\newblock Decentralised {LTL} monitoring.
\newblock In {\em Proc. of the 18th Int'l Symp. on Formal Methods (FM'12)},
  volume 7436 of {\em LNCS}, pages 85--100. Springer.

\bibitem[Carbone et~al., 2015]{carbone2015apache}
Carbone, P., Katsifodimos, A., Ewen, S., Markl, V., Haridi, S., and Tzoumas, K.
  (2015).
\newblock Apache {F}link: Stream and batch processing in a single engine.
\newblock {\em {IEEE} Data Eng. Bull.}, 38(4):28--38.

\bibitem[Convent et~al., 2018]{convent18tessla}
Convent, L., Hungerecker, S., Leucker, M., Scheffel, T., Schmitz, M., and
  Thoma, D. (2018).
\newblock {TeSSLa}: Temporal stream-based specification language.
\newblock In {\em Proc. of the 21th Brazilian Symp. on Formal Methods
  (SBMF'18)}, volume 11254 of {\em LNCS}, pages 144--162. Springer.

\bibitem[Cristian and Fetzer, 1999]{cristian99timed}
Cristian, F. and Fetzer, C. (1999).
\newblock The timed asynchronous distributed system model.
\newblock {\em IEEE Transactions on Parallel and Distributed Systems},
  10(6):642--657.

\bibitem[Cumin et~al., 2017]{cumin17orange4home}
Cumin, J., Lefebvre, G., Ramparany, F., and Crowley, J. (2017).
\newblock A dataset of routine daily activities in an instrumented home.
\newblock pages 413--425.

\bibitem[D'Angelo et~al., 2005]{dangelo05lola}
D'Angelo, B., Sankaranarayanan, S., S\'anchez, C., Robinson, W., Finkbeiner,
  B., Sipma, H.~B., Mehrotra, S., and Manna, Z. (2005).
\newblock {LOLA}: Runtime monitoring of synchronous systems.
\newblock In {\em Proc. of the 12th Int'l Symp. of Temporal Representation and
  Reasoning (TIME'05)}, pages 166--174. IEEE CS Press.

\bibitem[Danielsson and S{\'{a}}nchez, 2019]{danielsson19decentralized}
Danielsson, L.~M. and S{\'{a}}nchez, C. (2019).
\newblock Decentralized stream runtime verification.
\newblock In Finkbeiner, B. and Mariani, L., editors, {\em Runtime Verification
  - 19th International Conference, {RV} 2019, Porto, Portugal, October 8-11,
  2019, Proceedings}, volume 11757 of {\em Lecture Notes in Computer Science},
  pages 185--201. Springer.

\bibitem[Eisner et~al., 2003]{eisner03reasoning}
Eisner, C., Fisman, D., Havlicek, J., Lustig, Y., McIsaac, A., and Campenhout,
  D.~V. (2003).
\newblock Reasoning with temporal logic on truncated paths.
\newblock In {\em Proc. of the 15th Int'l Conf. on Computer Aided Verification
  (CAV'03)}, volume 2725 of {\em LNCS}, pages 27--39. Springer.

\bibitem[El-Hokayem and Falcone, 2017a]{elhokayem17monitoring}
El-Hokayem, A. and Falcone, Y. (2017a).
\newblock Monitoring decentralized specifications.
\newblock In {\em Proc. of the 26th ACM SIGSOFT Int'l Symp. on Software Testing
  and Analysis (ISSTA'17)}, pages 125--135. ACM.

\bibitem[El-Hokayem and Falcone, 2017b]{elhokayem17themis}
El-Hokayem, A. and Falcone, Y. (2017b).
\newblock {THEMIS: A Tool for Decentralized Monitoring Algorithms}.
\newblock In {\em Proc. of the 26th ACM SIGSOFT Int'l Symp. on Software Testing
  and Analysis (ISSTA'17)}, pages 125--135. ACM.

\bibitem[El-Hokayem and Falcone, 2020]{monitoringDecentSpecs2020El-Hokayem}
El-Hokayem, A. and Falcone, Y. (2020).
\newblock On the monitoring of decentralized specifications: Semantics,
  properties, analysis, and simulation.
\newblock {\em ACM Trans. Softw. Eng. Methodol.}, 29(1).

\bibitem[Faymonville et~al., 2016]{faymonville16stream}
Faymonville, P., Finkbeiner, B., Schirmer, S., and Torfah, H. (2016).
\newblock A stream-based specification language for network monitoring.
\newblock In {\em Proc. of the 16th Int'l Conf. on Runtime Verification
  (RV'16)}, volume 10012 of {\em LNCS}, pages 152--168. Springer.

\bibitem[Faymonville et~al., 2019]{faymonville19streamlab}
Faymonville, P., Finkbeiner, B., Schledjewski, M., Schwenger, M., Stenger, M.,
  Tentrup, L., and Hazem, T. (2019).
\newblock {StreamLAB}: Stream-based monitoring of cyber-physical systems.
\newblock In {\em Proc. of the 31st Int'l Conf. on Computer-Aided Verification
  (CAV'19)}, volume 11561 of {\em LNCS}, pages 421--431. Springer.

\bibitem[Francalanza et~al., 2018]{francalanza18decentralised}
Francalanza, A., P{\'{e}}rez, J.~A., and S{\'{a}}nchez, C. (2018).
\newblock Runtime verification for decentralised and distributed systems.
\newblock In Bartocci, E. and Falcone, Y., editors, {\em Lectures on Runtime
  Verification - Introductory and Advanced Topics}, volume 10457 of {\em
  Lecture Notes in Computer Science}, pages 176--210. Springer.

\bibitem[Ganguly et~al., 2021]{ganguly2021distributed}
Ganguly, R., Momtaz, A., and Bonakdarpour, B. (2021).
\newblock {Distributed Runtime Verification Under Partial Synchrony}.
\newblock In Bramas, Q., Oshman, R., and Romano, P., editors, {\em 24th
  International Conference on Principles of Distributed Systems (OPODIS 2020)},
  volume 184 of {\em Leibniz International Proceedings in Informatics
  (LIPIcs)}, pages 20:1--20:17, Dagstuhl, Germany. Schloss
  Dagstuhl--Leibniz-Zentrum f{\"u}r Informatik.

\bibitem[Gorostiaga et~al., 2020]{gorostiaga20unifying}
Gorostiaga, F., Danielsson, L.~M., and S{\'{a}}nchez, C. (2020).
\newblock Unifying the time-event spectrum for stream runtime verification.
\newblock In Deshmukh, J. and Nickovic, D., editors, {\em Runtime Verification
  - 20th International Conference, {RV} 2020, Los Angeles, CA, USA, October
  6-9, 2020, Proceedings}, volume 12399 of {\em Lecture Notes in Computer
  Science}, pages 462--481. Springer.

\bibitem[Gorostiaga and S\'anchez, 2018]{gorostiaga18striver}
Gorostiaga, F. and S\'anchez, C. (2018).
\newblock {S}triver: Stream runtime verification for real-time event-streams.
\newblock In {\em Proc. of the 18th Int'l Conf. on Runtime Verification
  (RV'18)}, volume 11237 of {\em LNCS}, pages 282--298. Springer.

\bibitem[Havelund and Ro\c{s}u, 2002]{havelund02synthesizing}
Havelund, K. and Ro\c{s}u, G. (2002).
\newblock Synthesizing monitors for safety properties.
\newblock In {\em Proc. of the 8th Int'l Conf. on Tools and Algorithms for the
  Construction and Analysis of Systems (TACAS'02)}, volume 2280 of {\em LNCS},
  pages 342--356. Springer-Verlag.

\bibitem[Jaber et~al., 2020]{globalChoreosSynthesis2020El-Hokayem}
Jaber, M., Falcone, Y., Attie, P., Khalil, A.-A., Hallal, R., and El-Hokayem,
  A. (2020).
\newblock From global choreographies to verifiable efficient distributed
  implementations.
\newblock {\em Journal of Logical and Algebraic Methods in Programming},
  115:100577.

\bibitem[Kaupp et~al., 2021]{kaupp2021contextDataset}
Kaupp, L., Webert, H., Nazemi, K., Humm, B., and Simons, S. (2021).
\newblock Context: An industry 4.0 dataset of contextual faults in a smart
  factory.
\newblock {\em Procedia Computer Science}, 180:492--501.
\newblock Proceedings of the 2nd International Conference on Industry 4.0 and
  Smart Manufacturing (ISM 2020).

\bibitem[Kazemlou and Bonakdarpour, 2018]{crash-resilient2018Borzoo}
Kazemlou, S. and Bonakdarpour, B. (2018).
\newblock Crash-resilient decentralized synchronous runtime verification.
\newblock In {\em 2018 IEEE 37th Symposium on Reliable Distributed Systems
  (SRDS)}, pages 207--212.

\bibitem[Pajuelo-Holguera et~al., 2020]{pajuelo-holguera2020smartPolitech}
Pajuelo-Holguera, F., Gómez-Pulido, J.~A., and Ortega, F. (2020).
\newblock Recommender systems for sensor-based ambient control in academic
  facilities.
\newblock {\em Engineering Applications of Artificial Intelligence}, 96:103993.

\bibitem[Perez et~al., 2020]{perez20copilot}
Perez, I., Dedden, F., and Goodloe, A. (2020).
\newblock Copilot 3.
\newblock Technical Report NASA/TM–2020–220587, NASA Langley Research
  Center.

\bibitem[Pike et~al., 2010]{pike10copilot}
Pike, L., Goodloe, A., Morisset, R., and Niller, S. (2010).
\newblock {Copilot}: A hard real-time runtime monitor.
\newblock In {\em Proc. of the 1st Int'l Conf. on Runtime Verification
  (RV'10)}, volume 6418 of {\em LNCS}, pages 345--359. Springer.

\bibitem[Pike et~al., 2013]{pike13copilot}
Pike, L., Wegmann, N., Niller, S., and Goodloe, A. (2013).
\newblock Copilot: monitoring embedded systems.
\newblock {\em Innovations in Systems and Software Engineering}, 9(4):235--255.

\bibitem[Quoc et~al., 2017]{dolequoc19PrivApprox}
Quoc, D.~L., Beck, M., Bhatotia, P., Chen, R., Fetzer, C., and Strufe, T.
  (2017).
\newblock Privapprox: Privacy-preserving stream analytics.
\newblock In {\em 2017 {USENIX} Annual Technical Conference ({USENIX} {ATC}
  17)}, pages 659--672, Santa Clara, CA. {USENIX} Association.

\bibitem[Ro\c{s}u and Havelund, 2005]{rosu05rewriting}
Ro\c{s}u, G. and Havelund, K. (2005).
\newblock Rewriting-based techniques for runtime verification.
\newblock {\em Automated Software Engineering}, 12(2):151--197.

\bibitem[S\'anchez, 2018]{sanchez18online}
S\'anchez, C. (2018).
\newblock Online and offline stream runtime verification of synchronous
  systems.
\newblock In {\em Proc. of the 18th Int'l Conf. on Runtime Verification
  (RV'18)}, volume 11237 of {\em LNCS}, pages 138--163. Springer.

\bibitem[Sen and Ro\c{s}u, 2003]{sen03generating}
Sen, K. and Ro\c{s}u, G. (2003).
\newblock Generating optimal monitors for extended regular expressions.
\newblock In Sokolsky, O. and Viswanathan, M., editors, {\em Electronic Notes
  in Theoretical Computer Science}, volume~89. Elsevier.

\bibitem[Sen et~al., 2004]{sen04efficient}
Sen, K., Vardhan, A., Agha, G., and Rosu, G. (2004).
\newblock Efficient decentralized monitoring of safety in distributed systems.
\newblock In {\em Proc. of the 26th Int'l Conf. on Software Engineering
  (ICSE'04)}, pages 418--427. IEEE CS Press.

\bibitem[Zorin and Stukach, 2020]{zorin2020tomskHeating}
Zorin, P. and Stukach, O. (2020).
\newblock Data of heating meters from residential buildings in tomsk (russia)
  for statistical modeling of the thermal characteristics of buildings.

\end{thebibliography}
%just for ieeeaccess
%\EOD 
\end{document}